\documentclass[sigconf,nonacm]{acmart}
\setcopyright{none}
\newenvironment{ai}{\section*{AI Use}}{}
\usepackage{url}
\usepackage{xcolor}

\usepackage[ruled,vlined,linesnumbered]{algorithm2e}
\makeatletter
\@ifpackageloaded{amssymb}{}{}
\makeatother
\usepackage{amsmath,amssymb,amsthm}
\usepackage{mathtools}
\usepackage{placeins}
\usepackage{booktabs}
\usepackage{multirow}
\usepackage{graphicx}
\usepackage{xcolor}
\usepackage{enumitem}
\usepackage{bm}
\usepackage{longtable}
\usepackage{subcaption}
\usepackage{mdframed}
\usepackage{placeins}
\usepackage{makecell}
\usepackage{tabularx}
\usepackage{float}
\usepackage{capt-of}
\usepackage{etoolbox}
\newcommand{\blacktables}{%
  \color{black}\colorlet{green}{black}%
  \captionsetup{font={color=black},labelfont={color=black}}}
\AtBeginEnvironment{table}{\blacktables}
\AtBeginEnvironment{table*}{\blacktables}
\newtheorem{theorem}{Theorem}[section]
\newtheorem{lemma}[theorem]{Lemma}
\newtheorem{proposition}[theorem]{Proposition}
\newtheorem{corollary}[theorem]{Corollary}
\theoremstyle{definition}
\newtheorem{definition}[theorem]{Definition}
\theoremstyle{remark}
\newtheorem{remark}[theorem]{Remark}
\newcommand{\Lap}{\mathrm{Lap}}
\newcommand{\Var}{\mathrm{Var}}
\newcommand{\E}{\mathbb{E}}
\newcommand{\R}{\mathbb{R}}
\newcommand{\eps}{\varepsilon}
\newcommand{\Bcell}[2]{B_{#1;\,#2}}
\newcommand{\tBcell}[2]{\tilde{B}_{#1;\,#2}}
\newcommand{\yb}{1-y}
\newcommand{\methodname}{PROBES}
\newcommand{\bk}{\bm{k}}
\newcommand{\best}[1]{\textbf{#1}}
\newcommand{\second}[1]{\underline{#1}}

\begin{document}
\raggedbottom

  \title[PROBES: A Bernstein Simplex Approach to Differentially Private OLS Estimation]{PROBES: A Bernstein Simplex Approach to Differentially Private OLS Estimation}

\author{Sasmita Harini S}
\authornote{Work done while an undergraduate student at the Indian Institute of Science.}
\affiliation{%
  \institution{Microsoft}
  \city{Bangalore}
  \state{}
  \country{India}}
\email{t-sasmitas@microsoft.com}

\author{Anshoo Tandon}
\affiliation{%
  \institution{CDPG, Indian Institute of Science}
  \city{Bangalore}
  \state{}
  \country{India}}
\email{anshoo.tandon@gmail.com}

\renewcommand{\shortauthors}{Sasmita Harini S and Anshoo Tandon}

 \begin{abstract}
  Differentially private ordinary least squares regression on bounded continuous data requires the joint release of multiple sufficient statistics.
For $p$ features per data sample, we have
$K=(p+1)(p+4)/2$ distinct statistics including intercept and count. Under add/remove adjacency and fixed
public bounds $[0,1]$ for features and response variables, the direct workload has
$\ell_1$-sensitivity $K$.
Releasing the direct workload as one Laplace vector at the full privacy budget
requires noise calibrated to its joint $\ell_1$-sensitivity $K$.
We introduce \methodname{} (\emph{\textbf{PR}ivate \textbf{O}rdinary least squares via \textbf{BE}rnstein \textbf{S}implex}), a pure $\varepsilon$-DP mechanism that exploits the common algebraic structure shared across the sufficient statistics, releases \emph{a single} Laplace vector at the full unsplit budget $\eps$, and recovers \emph{every} sufficient statistic for multivariate linear regression on $p$ bounded features by post-processing.
The construction rests on the Bernstein simplex embedding where each record is mapped to a non-negative
vector of dimension $3^p \times 2$ that sums to exactly $1$, giving
a global $\ell_1$-sensitivity of $1$ regardless of $p$.
We also present \textsc{Ada-PROBES}, which retains the single-shot Bernstein release of \textsc{PROBES} and recovers coefficients via a noise-calibrated ridge solve that stabilises matrix inversion at no extra privacy cost.
For larger feature dimensions, we replace the exponential-size tensor with $O(p^2)$ constant-size Steiner simplex blocks where suitable designs exist, retaining pure $\varepsilon$-DP. 
  Experiments on six real-world regression benchmarks show that
\textsc{Ada-PROBES} is within $1\%$ of non-private OLS on five of the six
datasets for $p\leq4$ at $\varepsilon=1$, and achieves lower test error
than the competing methods in most evaluated settings. The source code is available at
\url{https://github.com/sasmitahs/PROBES}.
 \end{abstract}


\sloppy
 \maketitle

\section{Introduction}
\label{sec:intro}

Modern data   analysis tools  increasingly act as the privacy boundary between
sensitive   user-data  records and downstream analytical workloads.  Linear regression
is widely used to analyse individual-level data, including settings in which
the records used for model training contain sensitive information whose privacy must be meaningfully protected.  Government
statistical agencies, health-data exchanges, and consumer-facing
analytics products  routinely serve regression coefficients, forecasts,
and policy estimates derived from records describing individual people,
vehicles, or transactions \cite{NIST2025,abowd2018us}. For these systems,
\emph{differential privacy}~\cite{dwork2006calibrating}  provides a rigorous,
composition-friendly bound on what an  analyst can learn about an  individual
record from the released output.

Among the  workloads serviced by such systems,   ordinary
least squares (OLS) regression on bounded continuous data remains
challenging under differential privacy because the estimator depends
on multiple second-order and cross-product statistics.  The estimator itself is closed-form
 $\hat{\bm\theta} = (X^\top X)^{-1} X^\top y$. The private analogue is
also conceptually simple, privatise the entries of $X^\top X$ and
$X^\top y$, invert, and return. Yet for $p$-dimensional multivariate
linear regression, the two ingredients --- the Gram matrix
$X^\top X$ and the cross-product vector $X^\top y$ --- contain
$K=\binom{p+2}{2}+(p+1)=\Theta(p^2)$ distinct sums. With public bounds
$x_{ij},y_i\in[0,1]$, fixed public preprocessing and add/remove adjacency,
the vector of unique entries has $\ell_1$-sensitivity $K$. A single joint Laplace release of
this vector uses independent $\Lap(K/\varepsilon)$ coordinates; its per-statistic
variance is $2K^2/\varepsilon^2=\Theta(p^4/\varepsilon^2)$.
This Laplace mechanism is just a baseline; the algebraic structure shared across sufficient statistics can be exploited via a
different representation.
\paragraph{Practical privacy relevance.} In this work, we present the
\methodname{} mechanism that provides
a single-shot $\varepsilon$-differentially private mechanism for
multivariate linear regression on bounded data, allowing the required
OLS sufficient statistics to be recovered from a single privatized vector
release. We evaluate the resulting privacy--utility trade-offs on six
real-world regression datasets, in addition to synthetic benchmark,
across varying feature dimensionality, sample size, and privacy budget. The superior performance of \textsc{PROBES}
for $p \le 4$ is demonstrated through extensive experimental results:
the estimation-error variances with \textsc{PROBES} are $\Theta(3^{p}/\varepsilon^{2})$,
compared with $\Theta(p^4/\varepsilon^2)$ for a direct joint
Laplace release. This difference arises from the representation and
joint sensitivity, not from making fewer releases.

 The sufficient statistics for multivariate linear regression are  coordinated moments of the \emph{same}
bounded records, and   the bounded domain $[0,1]^{p+1}$ allows these statistics to be represented through a common polynomial basis.  The key question   here  is:

\begin{quote}
\itshape   Can we exploit the shared algebraic structure of OLS sufficient statistics to recover them from a single Laplace release by linear post-processing?
\end{quote}

\noindent Our construction exploits a classical object from approximation theory using the Bernstein polynomial basis, and we provide
detailed analytical and numerical comparisons with competing private linear regression mechanisms for different settings.

\paragraph{The Bernstein simplex insight (linear regression).}
For multivariate linear regression with $p$ bounded features, we use the
Bernstein basis of order $2$ on each coordinate:
$p_0(x) = (1-x)^2$, $p_1(x) = 2x(1-x)$, $p_2(x) = x^2$. These three
polynomials are non-negative on $[0,1]$ and sum to $1$. Tensoring
across the $p$ input coordinates and once across the response
decomposition $\{y, 1{-}y\}$ gives an embedding
$\phi : [0,1]^{p+1} \to \Delta_{N-1}$ into the $(N-1)$-simplex with
$N = 3^p \times 2$, and \emph{every} record maps to an $\ell_1$-unit
vector. The Bernstein representation is useful because every record
with features and response in $[0,1]$ is mapped to a non-negative
vector whose entries sum to $1$. Consequently, under add/remove
adjacency, the summed embedding has global $\ell_1$-sensitivity
exactly $1$, regardless of the data values or the number of features. A single Laplace vector release at scale $1/\varepsilon$ then suffices for
$\varepsilon$-DP.

\paragraph{Practical dimensionality.}
The cell count $3^{p}\times 2$ and the estimation-error variance
$\Theta(3^{p}/\varepsilon^{2})$ both grow exponentially in $p$,
limiting \methodname{} to   $p \le 6$  in practice.
S-PROBES, introduced in Section~\ref{sec:steiner_tmlr_main}, extends PROBES to larger feature dimensions via Steiner system based decomposition of features into smaller \emph{blocks}, and is evaluated up to $p=24$.
This restriction is motivated by the broader challenge of private regression in relatively
higher-dimensional settings. Dick et al.~\cite{dick2023better} show that the utility of
existing ``plug-and-play'' differentially private linear regression methods can deteriorate
as the feature dimension increases, and demonstrate that a differentially private
feature-selection step can substantially broaden their applicability. PROBES
addresses the regression step once a low-dimensional set of features has been specified. We remark that we do
not address how the feature subset itself is selected privately; incorporating a private
feature-selection procedure is complementary to PROBES and lies outside the scope of the
present work.

Our main contributions are as follows:
\begin{enumerate}[leftmargin=*]
\item \textbf{The Bernstein simplex embedding
for multivariate regression
(\S\ref{sec:bernstein}--\S\ref{sec:PROBES}).}
We develop a Bernstein simplex embedding of bounded records for multivariate regression
whose summed representation has global $\ell_1$-sensitivity
exactly $1$ under add/remove adjacency. Building on this
embedding, we develop \methodname{}, a pure $\varepsilon$-DP
mechanism that recovers all multivariate OLS sufficient
statistics by linear post-processing of a single Laplace
vector release.

\item \textbf{Empirical evaluation (\S\ref{sec:experiments}).}
  A systematic comparison on six real-world benchmarks and one
synthetic dataset against \textsc{AdaSSP}, \textsc{BinAgg}, and
\textsc{DiffPrivLib}.
Throughout, \textsc{Ada-PROBES} denotes \textsc{PROBES} followed by the noise-aware regularised solve of~\eqref{eq:ridge_solve}; it differs from base \textsc{PROBES} only in this deterministic post-processing step.

\item \textbf{The S-PROBES mechanism (\S\ref{sec:steiner_tmlr_main}).}
 A Steiner system based extension of PROBES that retains pure $\varepsilon$-DP and scales well with increasing $p$, is evaluated up to $p=24$. Its performance varies
with the dataset, privacy budget, and feature dimension and is
competitive in several evaluated settings.

\end{enumerate}

\paragraph{Roadmap.}
\S\ref{sec:related}--\S\ref{sec:prelim} cover related work and background.
\S\ref{sec:bernstein} develops the Bernstein polynomial foundations.
\S\ref{sec:PROBES} presents the \methodname{} mechanism.
\S\ref{sec:experiments} reports empirical results. \S\ref{sec:steiner_tmlr_main} describes S-PROBES, our Steiner system based polynomial-size variant of PROBES. \S\ref{sec:conclusion} concludes. Appendix~\ref{proof:bernstein} contains proofs.
Appendix~\ref{sec:mechanism} gives full algorithmic details. Appendix~\ref{sec:steiner_tmlr_appendix} discusses the details of S-PROBES mechanism. Appendix~\ref{sec:error_revision} states conditional coefficient bounds.

\section{Related Work}
\label{sec:related}

  Differentially  private linear regression   has been studied through several
distinct  perturbation paradigms.  A first class of methods privatizes the
optimization procedure itself.  \textbf{Objective perturbation} modifies the
loss function before optimization by injecting a random linear term, ensuring
 that  the resulting minimizer satisfies   differential privacy .
\citet{kifer2012private} established tight error bounds for private convex
empirical risk minimization in this framework, while
\citet{zhang2012functional} proposed the \emph{Functional Mechanism}, which
 instead  perturbs the polynomial coefficients of the   objective function and
applies naturally to regression and other optimization-based analyses.

A second prominent approach is  \textbf{Differentially Private Stochastic
Gradient Descent (DP-SGD)}  , which  clips per-sample gradients and injects
calibrated   noise during iterative optimization
\citep{bassily2014private,abadi2016deep,cai2021cost}.
These works provide foundational privacy--utility guarantees for private
optimization, while in practice  DP-SGD   typically requires careful choices of
clipping thresholds, noise levels, and other optimization hyperparameters.
Recent audits have also highlighted that privacy-critical implementations
require careful verification, with  several open-source   systems found to
 overstate their effective privacy guarantees
  \citep{annamalai2024shuffle,chua2024howprivate}.

For OLS, however, a particularly natural alternative is to
privatize the quantities on which the estimator directly depends. Under
\textbf{sufficient-statistic perturbation (SSP)}, the OLS solution is computed
from privatized versions of
$(\mathbf{X}^{\top}\mathbf{X},\mathbf{X}^{\top}\mathbf{y})$ rather than by
perturbing the optimization process itself.
\citet{dwork2014algorithmic} established foundational
 sensitivity-based   mechanisms for releasing such statistics , and
\citet{sheffet2017differentially}   developed private OLS procedures with
statistical  guarantees. Recognizing that   straightforward SSP can depend
strongly on assumptions about  the feature domain ,
\citet{wang2018PrivateLinearRegression} proposed  the adaptive methods
 \emph{AdaSSP} and \emph{AdaOPS},   which privately estimate quantities needed to
calibrate the perturbation and thereby obtain strong utility guarantees under
weaker prior domain information.

Azize and Basu~\citep{azize2024concentrated} study zero concentrated
differential privacy for bandits under replacement adjacency on reward inputs. Their AdaC-OFUL algorithm adds Gaussian noise to the reward-weighted action sum; the design matrix
$V_t=\lambda I+\sum_{s=1}^t a_s a_s^\top$, corresponding to a regularized
$X^\top X$, is not directly perturbed. This differs from our pure
$\varepsilon$-DP setting under add/remove adjacency, where both covariates
as well as the response variables are private.

  Another complementary perspective on private sufficient-statistic release
is provided by the matrix mechanism~\cite{li2015matrix}, which answers a
workload of linear queries by privately measuring a suitably chosen collection
of strategy queries and reconstructing the desired workload from their noisy
answers. The course
materials of Honaker and Vadhan~\cite{vadhan_matrix_lecture} illustrate the
matrix mechanism using a simple linear-regression coefficient reconstructed
from privately answered sufficient-statistic queries. More generally, the
entries of $X^\top X$ and $X^\top y$ can be viewed as a collection of aggregate
queries, making the matrix-mechanism framework applicable to their private
release. PROBES instead exploits algebraic structure arising from the
Bernstein-simplex representation in constructing its private estimates, rather
than selecting a workload-specific query strategy. Thus, the two approaches
exploit different forms of structure and are not mutually exclusive.

Beyond optimization-based and sufficient-statistic approaches , several works
  develop structurally different mechanisms for private regression.
\citet{alabi2022differentially} studied differentially private \emph{simple}
 linear regression and proposed \emph{DP-Theil--Sen}, a   privatized
 nonparametric slope estimator based on   pairwise slopes. In a different
direction,  \citet{lin2025dplrsynthetic} introduced the \emph{BinAgg}
framework, which discretizes the feature space , privately aggregates
  statistics within bins , and reconstructs regression estimates   while  also
supporting downstream   synthetic-data generation. These approaches illustrate
that private regression need not rely exclusively on perturbing the classical
OLS optimization problem or its standard sufficient statistics.

Another  line of work   exploits special algebraic structure  in the add-remove
model of   differential privacy . \citet{kulesza2024} and
\citet{fitzsimons2024}   study phenomena in which information about quantities
such as counts or means can be obtained with reduced or effectively ``free''
privacy cost through structured transformations and joint private estimation .
Building on this line of work, \citet{harini2026refined} proposed
\emph{DP-RSS}.  It releases two groups of complementary simplex
statistics with Laplace noise and combines independent unbiased estimates
of the same sufficient statistic using inverse-variance weights.
This line of work is particularly close in spirit
to PROBES in that it seeks improved utility by exploiting algebraic structure
among related statistics, rather than treating every target quantity as an
independent query.

Bernstein polynomial structure has also appeared in differential privacy,
although in settings different from private OLS sufficient-statistic release.
\citet{alda2017bernstein} introduced the \emph{Bernstein Mechanism} for
differentially private function release, using Bernstein polynomial
approximation and perturbation to enable the release of an entire function.
Independently, and concurrently with our work, Takakura et
al.~\cite{takakura2025optimalvariancecovarianceestimation} developed the
B\'{e}zier mechanism, which leverages Bernstein/B\'{e}zier polynomial
structure for private variance and covariance estimation in the add-remove
model.

Takakura et al. also mention potential extension of their construction to multivariate tensor
products. In particular, $\{1-y,y\}$ is a standard degree-one Bernstein
response basis, not a novel decomposition. In our work, we specify the full OLS
moment workload, its recovery covariances, and a regularized solve for multivariate regression; we do
not claim optimal variance factors among private mechanisms or priority
over the shared Bernstein construction.

Finally, practical deployment depends not only on the underlying mechanism but
also on its implementation. The  \texttt{diffprivlib} library
\citep{holohan2019diffprivlib} provides   widely used
 scikit-learn-compatible   tools for differentially private machine learning,
including regression. Recent work identifying and correcting an  inconsistency
in its regression implementation  ~\citep{cebere2026bug} further illustrates
 the importance of formal auditing  and implementation-level verification  for
privacy-critical  statistical  software.

\section{Preliminaries}
\label{sec:prelim}

\subsection{Notation}
We write $D = \{(\bm{x}_i, y_i)\}_{i=1}^n$ for a dataset of $n$ records
with $p$ bounded features $x_{i,j} \in [0,1]$, $1 \le j \le p$, and response $y_i \in [0,1]$,
so that $\bm{x}_i \in [0,1]^p$ is the input vector of record $i$.
The full design matrix is $X \in \mathbb{R}^{n \times (p+1)}$ and
$\bm{\theta} = (\beta_0, \beta_1, \ldots, \beta_p)^\top \in \mathbb{R}^{p+1}$ denotes the OLS coefficient vector.
Two datasets $D \sim D'$ denote add-remove neighbours.
We write $S_{\bm{x}^{\bm{m}}}$ and $S_{\bm{x}^{\bm{m}}y}$ for the monomial sums that serve as sufficient statistics,
with $S$ and $\hat{S}$ denoting the true and private sufficient statistic respectively;
  the Bernstein cell indexed by ${k}$ and response component $r$ is
the product of the coordinate-wise Bernstein basis values and $r$'s
response weight (either $y$ or $1-y$). Its sum over records is
$\Bcell{{k}}{r}$,  and $\tBcell{{k}}{r}$   is its noisy release;
 $p_k(x)$ is the order-$2d$ Bernstein basis polynomial, $w_{m,k}$ is the Bernstein monomial recovery weight,
and $\sigma_{m,d}^2$ is the Bernstein variance factor.
We write $\Lap(b)$ for the zero-mean Laplace distribution, scale $b$, variance $2b^2$,
and $\varepsilon$ for the privacy budget under pure $\varepsilon$-DP.
For a vector-valued query $f$, the global $\ell_1$-sensitivity is
$\Delta_1(f) = \sup_{D \sim D'} \|f(D) - f(D')\|_1$.
Finally, $\rho$ denotes the analytical noise standard deviation used in regularization,
and $\lambda_{\min}(\cdot)$ denotes the smallest eigenvalue.

\subsection{Neighbouring Datasets and Differential Privacy}

We work throughout in the \emph{add-remove}
model of neighbouring datasets, which is the standard model for pure
$\varepsilon$-DP and is the model adopted by the free-lunch constructions
of~\citet{kulesza2024} and~\citet{fitzsimons2024} that motivate our work.

\begin{definition}[Dataset and add-remove neighbours]
\label{def:neighbours}
A \emph{dataset} is a multiset $D = \{z_1, \ldots, z_n\} \subset \mathcal{Z}$
of records drawn from a universe $\mathcal{Z}$.
Two datasets $D, D' \subset \mathcal{Z}$ are \emph{add-remove neighbours},
written $D \sim D'$, if their symmetric difference satisfies
$|D \,\triangle\, D'| = 1$, i.e.\ $D'$ is obtained from $D$ by the
insertion or deletion of exactly one record.
\end{definition}

\begin{definition}[$\varepsilon$-Differential Privacy~\citep{dwork2006calibrating}]
\label{def:dp}
Let $\varepsilon \geq 0$.
A randomised mechanism $\mathcal{M} : \mathcal{Z}^* \to \mathcal{Y}$
satisfies \emph{pure $\varepsilon$-differential privacy} ($\varepsilon$-DP)
if, for every pair of add-remove neighbours $D \sim D'$ and for every
measurable event $S \subseteq \mathcal{Y}$,
\begin{equation}
    \Pr\bigl[\mathcal{M}(D) \in S\bigr]
    \;\leq\;
    e^{\varepsilon}\,\Pr\bigl[\mathcal{M}(D') \in S\bigr].
    \label{eq:dp}
\end{equation}
The parameter $\varepsilon > 0$ is the \emph{privacy budget}; smaller values
yield stronger privacy guarantees.
\end{definition}

\noindent
The central quantity governing the noise required by any DP mechanism is
the \emph{sensitivity} of the query being privatised.

\begin{definition}[Global $\ell_1$-sensitivity]
\label{def:sensitivity}
The \emph{global $\ell_1$-sensitivity} of a function
$f : \mathcal{Z}^* \to \mathbb{R}^k$ is
\begin{equation}
    \Delta_1(f)
    \;=\;
    \sup_{\substack{D,\,D'\,:\,D \sim D'}}
    \bigl\|f(D) - f(D')\bigr\|_1.
\end{equation}
\end{definition}

\noindent
We collect the three standard properties of differential privacy that are
used throughout the paper.

\begin{proposition}[Standard DP properties~\citep{dwork2006calibrating,dwork2014algorithmic}]
\label{prop:dp_facts}
Let $f : \mathcal{Z}^* \to \mathbb{R}^k$ with $\Delta_1(f) < \infty$,
let $\mathcal{M}$ be an $\varepsilon$-DP mechanism, and let
$g : \mathcal{Y} \to \mathcal{Y}'$ be an arbitrary (possibly randomised)
function.

\begin{enumerate}[label=(\roman*)]

\item \textbf{Laplace mechanism.}
Define $\mathcal{M}_{\mathrm{Lap}}(D) = f(D) + \boldsymbol{Z}$ where
$\boldsymbol{Z} = (Z_1,\ldots,Z_k)$ with
$Z_i \overset{\mathrm{i.i.d.}}{\sim} \mathrm{Lap}(0,\,\Delta_1(f)/\varepsilon)$.
Then $\mathcal{M}_{\mathrm{Lap}}$ satisfies $\varepsilon$-DP.
Formally, the Laplace distribution with scale $b > 0$ has density
$p(z) = \tfrac{1}{2b}\exp(-|z|/b)$, so the injected noise has standard
deviation $\sigma = b\sqrt{2} = \sqrt{2}\,\Delta_1(f)/\varepsilon$.

\item \textbf{Post-processing immunity.}
For any (possibly randomised) function
$g : \mathcal{Y} \to \mathcal{Y}'$, the composed mechanism
$g \circ \mathcal{M}$ satisfies $\varepsilon$-DP.
In particular, no amount of data-independent post-processing can
degrade a privacy guarantee.

\item \textbf{Sequential composition.}
If $\mathcal{M}_1$ satisfies $\varepsilon_1$-DP and $\mathcal{M}_2$
satisfies $\varepsilon_2$-DP, and both are evaluated on the same dataset
(or $\mathcal{M}_2$ may depend on the output of $\mathcal{M}_1$), then
the joint release $(\mathcal{M}_1(D), \mathcal{M}_2(D))$ satisfies
$(\varepsilon_1 + \varepsilon_2)$-DP.
More generally, a sequence of $k$ adaptive mechanisms with budgets
$\varepsilon_1, \ldots, \varepsilon_k$ composes to
$\bigl(\sum_{i=1}^{k}\varepsilon_i\bigr)$-DP.

\end{enumerate}
\end{proposition}

\subsection{OLS Sufficient Statistics}
\label{subsec:ols_stats}
The \emph{Gram matrix} of a design matrix $X$ is the matrix of
pairwise inner products of its columns:
$G=X^\top X$, with $G_{st}=\langle X_{:s},X_{:t}\rangle$,
where $X_{:s}$ denotes column $s$ of $X$. It is symmetric and positive
semidefinite.

Ordinary least-squares (OLS) regression finds the coefficient vector
$\hat{\bm\theta}=(\hat\beta_0,\hat\beta_1,\ldots,\hat\beta_p)^\top$
minimising the squared prediction error
\begin{equation}
  \hat{\bm\theta}
  \;=\;
  \operatorname*{arg\,min}_{\bm\theta\in\mathbb{R}^{p+1}}
  \bigl\|\bm{y} - X\bm{\theta}\bigr\|_2^{2},
  \label{eq:ols_obj}
\end{equation}
where the \emph{augmented} design matrix $X\in\mathbb{R}^{n\times(p+1)}$
has $i$-th row $[1,\,x_{i1},\ldots,x_{ip}]$ (a leading $1$ for the
intercept $\beta_0$).

Differentiating~\eqref{eq:ols_obj} with respect to $\bm\theta$ and
setting the gradient to zero gives the \emph{normal equations}:
\begin{equation}
  \underbrace{(X^{\top}X)}_{\in\mathbb{R}^{(p+1)\times(p+1)}}
  \hat{\bm\theta}
  \;=\;
  \underbrace{X^{\top}\bm{y}}_{\in\mathbb{R}^{p+1}},
  \qquad
  \hat{\bm\theta}
  = \bigl(X^{\top}X\bigr)^{-1}X^{\top}\bm{y},
  \label{eq:normal}
\end{equation}
where the last equality holds whenever $X^{\top}X$ is positive definite.

\paragraph{Decomposing $X^{\top}X$ and $X^{\top}\bm{y}$.}
Write $x_{i0}:=1$ for all $i$.
The $(s,t)$-entry of the symmetric Gram matrix
($0\le s\le t\le p$) and the $s$-th entry of the cross-product
vector ($0\le s\le p$) are, respectively,
\begin{align}
  (X^{\top}X)_{st}
  &= \sum_{i=1}^{n}x_{is}\,x_{it}
   =: S_{x_{s}x_{t}},
   \label{eq:gram_entry}\\
  (X^{\top}\bm{y})_{s}
  &= \sum_{i=1}^{n}x_{is}\,y_{i}
   =: S_{x_{s}y}.
   \label{eq:cross_entry}
\end{align}
The corner cases are $S_{x_0x_0}=n$ (record count) and
$S_{x_0y}=S_y$ (response sum).

Expanding~\eqref{eq:gram_entry}--\eqref{eq:cross_entry}, the complete
set of scalar quantities needed to solve~\eqref{eq:normal} is
\begin{equation}
  \begin{aligned}
  \underbrace{n,\;
  \{S_{x_s}\}_{s=1}^{p},\;
  \{S_{x_s^2}\}_{s=1}^{p},\;
  \{S_{x_sx_t}\}_{1\le s<t\le p}
  }_{\frac{(p+1)(p+2)}{2}\text{ entries of }X^{\top}X} &\\
  {}\cup\;
  \underbrace{S_y,\;\{S_{x_sy}\}_{s=1}^{p}
  }_{(p+1)\text{ entries of }X^{\top}\bm{y}} &,
  \end{aligned}
  \label{eq:suff_stats}
\end{equation}
a total of
\begin{equation}
  \frac{(p+1)(p+2)}{2}+(p+1)
  \;=\;\binom{p+2}{2}+(p+1)
  \;=\;\Theta(p^{2})
  \label{eq:stat_count}
\end{equation}
distinct scalar sums.
We write $S$ for any true sufficient statistic in~\eqref{eq:suff_stats}
and $\hat{S}$ for its differentially private, post-processed estimate.

Each sum in~\eqref{eq:suff_stats} has a per-record contribution
in $[0,1]$ and individual add/remove sensitivity $1$. Stacking all
$K=(p+1)(p+4)/2$ unique sums gives joint sensitivities
$\Delta_1=K$ and $\Delta_2=\sqrt K$, attained by the all-ones record.
These bounds assume fixed public feature selection and preprocessing,
the domain $[0,1]^{p+1}$, and inclusion of the private count.
One joint release of this full vector uses $\Delta_1$ for Laplace
noise or $\Delta_2$ for Gaussian noise. The resulting joint Laplace
variance is $\Theta(p^4/\varepsilon^2)$ per
statistic at the full budget $\varepsilon$.

\paragraph{The \methodname{} insight.}
The quantities in~\eqref{eq:suff_stats} are \emph{not}
independent: they share a common algebraic structure rooted in the
bounded domain $[0,1]^{p+1}$.  \methodname{} exploits this structure
via a simplex embedding whose mechanism and privacy guarantee are
developed in full in \S\ref{sec:PROBES}.

\paragraph{Ridge regularisation in the private setting.}
Even when the true Gram matrix $X^\top X$ is invertible, its
\emph{privatised} counterpart $\widetilde{X^\top X} = X^\top X + E$
(where $E$ is the symmetric noise matrix induced by the Laplace
mechanism) may be ill-conditioned or indefinite with non-negligible
probability, particularly when the
privacy budget $\varepsilon$ is small. Directly inverting
$\widetilde{X^\top X}$ in~\eqref{eq:normal} can be numerically unstable and can produce very large coefficient errors when the noisy Gram matrix has eigenvalues close to zero. Ridge regression addresses this by
replacing~\eqref{eq:normal} with the regularised normal equations
\begin{equation}
  \hat{\bm\theta}_\lambda
  \;=\;
  \bigl(\widetilde{X^\top X} + \lambda I\bigr)^{-1}
  \widetilde{X^\top \bm{y}},
  \label{eq:ridge}
\end{equation}
where $\lambda > 0$ is a regularisation parameter.  Adding $\lambda I$
  does not by itself ensure positive definiteness when the noisy matrix is
indefinite; the spectral correction in~\eqref{eq:ridge_solve} first shifts
its smallest eigenvalue to zero .  Intuitively, ridge shrinks the coefficient
estimate towards zero, trading a small increase in bias for a
potentially large reduction in variance---a favourable bias--variance
tradeoff that becomes especially pronounced when the noisy Gram matrix
has small eigenvalues amplified by inversion.  In the differentially
private context, both \citet{sheffet2017differentially} and
\citet{wang2018PrivateLinearRegression} advocate choosing $\lambda$ proportional
to the noise scale, ensuring that the regularisation absorbs the dominant
source of instability without excessively shrinking signal directions.

\section{Bernstein Polynomial Foundations}
\label{sec:bernstein}

\subsection{Bernstein Polynomial Basis and Monomial Recovery}
\label{subsec:bernstein}

A key property of Bernstein basis polynomials of order $2d$ over
$[0,1]$ is that they form a \emph{partition of unity}: they are
non-negative everywhere on $[0,1]$ and sum pointwise to $1$.  This
means that for any $x \in [0,1]$, the vector
$(p_0(x), p_1(x), \ldots, p_{2d}(x))$ lies on the standard
$2d$-simplex $\Delta^{2d}$.  Its $\ell_1$-norm is therefore exactly
$1$, independent of $x$.  This bounded, fixed-norm structure is
precisely what enables a single Laplace release with a
\emph{data-independent} sensitivity.

\begin{definition}[Order-$2d$ Bernstein basis]
\label{def:bernstein_basis}

For $x \in [0,1]$ and non-negative integer $k \le 2d$, the
\emph{$k$-th Bernstein basis polynomial of order $2d$} is
\begin{equation}
  p_k(x)
  \;:=\;
  \binom{2d}{k} x^k (1-x)^{2d-k},
  \qquad 0 \le k \le 2d.
  \label{eq:bernstein_basis}
\end{equation}
Each $p_k$ is non-negative on $[0,1]$ and the family
$\{p_k\}_{k=0}^{2d}$ satisfies the \emph{partition-of-unity} property:
\begin{equation}
  \sum_{k=0}^{2d} p_k(x) \;=\; 1
  \qquad \forall\, x \in [0,1].
  \label{eq:partition_unity}
\end{equation}
\end{definition}

\noindent
The partition-of-unity property~\eqref{eq:partition_unity} follows
immediately from the binomial theorem: $\sum_{k=0}^{2d}\binom{2d}{k}
x^k(1-x)^{2d-k} = (x + (1-x))^{2d} = 1$.

\paragraph{The order-$2$ case ($d=1$).}
Setting $2d = 2$ gives the three basis polynomials
\begin{equation}
  p_0(x) = (1-x)^2,
  \qquad
  p_1(x) = 2x(1-x),
  \qquad
  p_2(x) = x^2,
  \label{eq:bernstein_d1}
\end{equation}
which satisfy $p_0(x)+p_1(x)+p_2(x)=1$ for all $x\in[0,1]$.
This is the base case used throughout \methodname{} for univariate features; larger values of $d$ support exact recovery of higher-degree monomials at the cost of a larger simplex dimension.

\begin{lemma}[Bernstein monomial recovery]
\label{lem:bernstein}

For any integer $0 \le m \le 2d$ and any $x \in [0,1]$,
\begin{equation}
  x^m
  \;=\;
  \sum_{k=m}^{2d}
  w_{m,k}\, p_k(x),
  \qquad
  w_{m,k}
  \;:=\;
  \frac{\displaystyle\binom{k}{m}}{\displaystyle\binom{2d}{m}},
  \label{eq:monomial_recovery}
\end{equation}
where the weights $w_{m,k}$ are non-negative and depend only on $m$,
$k$, and the order $2d$.
\end{lemma}

\begin{proof}
See Appendix~\ref{proof:bernstein}.
\end{proof}

\noindent
Lemma~\ref{lem:bernstein} is the critical algebraic fact underlying
\methodname{}: every monomial $x^m$ is a \emph{fixed, data-independent}
linear combination of the Bernstein basis values $p_k(x)$.
Consequently, if we can release the per-feature cell sums
$\tilde{B}_{k,j} := \sum_{i=1}^n p_k(x_{i,j})$ privately for each
feature $j \in \{1,\ldots,p\}$ and each $k \in \{0,\ldots,2d\}$,
we can recover any marginal monomial sum
$S_{x_j^m} = \sum_i x_{i,j}^m$ by the \emph{same} linear combination
applied to the released sums, with no additional privacy cost.
In the multivariate setting (\S\ref{sec:PROBES}), these per-feature
cell sums are further tensored across features and the response,
yielding the Bernstein cell sums $\Bcell{\bm{k}}{r}$.

For distinct features $s$ and $t$, where $s<t$ and integers $0\le a,b\le2d$,
applying Lemma~\ref{lem:bernstein} to both coordinates gives
\begin{equation}
\begingroup\color{black}
\begin{aligned}
 x_s^a x_t^b
 &= \left(\sum_{k=a}^{2d}w_{a,k}p_k(x_s)\right)
    \left(\sum_{\ell=b}^{2d}w_{b,\ell}p_\ell(x_t)\right)\\
 &= \sum_{k=a}^{2d}\sum_{\ell=b}^{2d}
       w_{a,k}w_{b,\ell}p_k(x_s)p_\ell(x_t).
\end{aligned}
\endgroup
\label{eq:cross_bernstein_identity}
\end{equation}

\paragraph{Explicit recovery at order $2d=2$.}
Applying~\eqref{eq:monomial_recovery} with $d=1$ gives $1 = p_0(x) + p_1(x) + p_2(x)$, $x = \tfrac{1}{2}\,p_1(x) + p_2(x)$, and $x^2 = p_2(x)$. These follow directly from substituting the weights: for $m=1$, $w_{1,1}=\binom{1}{1}/\binom{2}{1}=\tfrac{1}{2}$ and $w_{1,2}=\binom{2}{1}/\binom{2}{1}=1$; for $m=2$, $w_{2,2}=\binom{2}{2}/\binom{2}{2}=1$. A complete algorithm for the simple case $p=1$ and $d=1$ is provided in Appendix~\ref{sec:mechanism} (Algorithm~\ref{alg:dp_bsm_a}).

\section{The \methodname{} Mechanism}
\label{sec:PROBES}

We now describe \methodname{} (\emph{\textbf{PR}ivate \textbf{O}rdinary
least squares via \textbf{BE}rnstein \textbf{S}implex}),
a single-shot $\varepsilon$-DP mechanism for linear regression that releases
all sufficient statistics jointly under a single Laplace perturbation.
The key insight is a carefully chosen embedding of each record into a
simplex of fixed $\ell_1$-norm, which yields global $\ell_1$-sensitivity exactly equal to $1$
regardless of the number of features $p$.

\paragraph{Record embedding.}
Given a dataset $D = \{(\bm x_i, y_i)\}_{i=1}^n \subset [0,1]^{p+1}$,
we map each record to a vector of \emph{Bernstein cell values}.
For a multi-index $\bm k = (k_1,\ldots,k_p) \in \{0,1,2\}^p$ and a
response indicator $r \in \{y, \bar y\}$, define
\begin{equation}
  \phi(\bm x_i, y_i)_{(\bm k,\, r)}
  \;:=\;
  \Bigl(\prod_{s=1}^{p} p_{k_s}(x_{is})\Bigr)
  \cdot c_r(y_i),
  \label{eq:embedding}
\end{equation}
where $c_y(y) = y$ and $c_{\bar y}(y) = 1-y$.
We refer to each component $\phi(\bm{x}_i, y_i)_{(\bm{k},\, r)}$ as a
\emph{Bernstein cell}; intuitively, a cell is the weighted contribution
of a single record to one coordinate of the simplex, indexed by a
feature multi-index $\bm{k}$ and a response indicator $r$.
Since $\{p_0, p_1, p_2\}$ is a partition of unity on $[0,1]$
(Definition~\ref{def:bernstein_basis}), the product
$\prod_s p_{k_s}(x_{is})$ sums to $1$ over all $\bm k$.
Together with $c_y(y) + c_{\bar y}(y) = 1$, this gives
\begin{equation}
  \sum_{\bm k,\, r}
  \phi(\bm x_i, y_i)_{(\bm k,\, r)}
  \;=\; 1
  \qquad \forall\, (\bm x_i, y_i) \in [0,1]^{p+1},
  \label{eq:unit_norm}
\end{equation}
so $\phi(\bm x_i, y_i)$ always lies on the standard simplex
$\Delta^{N-1}$ with $N = 3^p \cdot 2$ coordinates
(see Table~\ref{tab:cell_counts} for the growth of $N$ with $p$).
The dataset-level \emph{Bernstein cell sums} are
$\Bcell{\bm k}{r} := \sum_{i=1}^n \phi(\bm x_i, y_i)_{(\bm k, r)}$.

\begingroup\color{black}
\begin{table}[t]
\centering
\caption{Number of Bernstein cells $N = 3^p \cdot 2$ as a function of
the number of features $p$. Despite exponential growth, the per-cell
noise scale is always $1/\varepsilon$ regardless of $p$.}
\label{tab:cell_counts}
\begin{tabular}{c|ccccc}
\toprule
$p$ & $1$ & $2$ & $3$ & $4$ & $5$ \\
\midrule
$N$ & $6$ & $18$ & $54$ & $162$ & $486$ \\
\bottomrule
\end{tabular}
\end{table}\endgroup

\paragraph{Unit sensitivity and single-shot release.}
Because every record contributes exactly $1$ in $\ell_1$-norm to the
cell vector~\eqref{eq:embedding}, adding or removing any single record
changes the cell sums by at most $1$ in $\ell_1$-norm, i.e.\
$\Delta_1(f) = 1$ under add-remove adjacency, independently of $p$.
This fixed, data-independent sensitivity is the mechanism's central
advantage: we draw $Z_{\bm k, r} \overset{\text{iid}}{\sim}
\mathrm{Lap}(1/\varepsilon)$ once and release
\begin{equation}
  \tBcell{\bm k}{r}
  \;:=\;
  \Bcell{\bm k}{r} + Z_{\bm k, r}
  \qquad \forall\, (\bm k, r).
  \label{eq:release}
\end{equation}
By the Laplace mechanism (Proposition~\ref{prop:dp_facts}), the
entire released vector $(\tBcell{\bm k}{r})_{\bm k, r}$ satisfies
$\varepsilon$-DP via a \emph{single} mechanism invocation, with no
composition and no budget splitting across statistics.
The full procedure is summarised in Algorithm~\ref{alg:PROBES}.

\paragraph{Variance of estimation error.}
Because each cell sum $\tBcell{\bm k}{r}$ is perturbed by an
independent $\mathrm{Lap}(1/\varepsilon)$ draw, the noise
amplification incurred when recovering a monomial sum from the
released cells is governed by the squared recovery weights.  We
formalise this as follows.

\begin{definition}[Variance factor]
\label{def:variance_factor}
For integers $0 \le m \le 2d$, the \emph{variance factor} is
\begin{equation}
  \sigma_{m,d}^2
  \;:=\;
  \sum_{k=m}^{2d} w_{m,k}^2
  \;=\;
  \sum_{k=m}^{2d}
  \frac{\binom{k}{m}^2}{\binom{2d}{m}^2}.
  \label{eq:var_factor}
\end{equation}
The second equality substitutes the recovery weights
$w_{m,k}$ from Lemma~\ref{lem:bernstein}.
\end{definition}

At order $2d=2$, the nonzero recovery weights are
$w_{0,0}=w_{0,1}=w_{0,2}=1$, $w_{1,1}=\tfrac12$,
$w_{1,2}=1$, and $w_{2,2}=1$. Substitution into~\eqref{eq:var_factor} gives
\begin{align*}
  \sigma_{0,1}^2 &= 1^2+1^2+1^2=3,\\
  \sigma_{1,1}^2 &= \bigl(\tfrac12\bigr)^2+1^2=\tfrac54,\\
  \sigma_{2,1}^2 &= 1^2=1.
\end{align*}
At order $2$, higher-degree monomials use fewer recovery weights and therefore incur less noise amplification in this representation while
the zero-th moment ($m=0$, the count) accumulates noise from all
three basis functions.  The factor $\sigma_{2,1}^2=1$ reflects the
fact that $x^2 = p_2(x)$ is itself a Bernstein basis element at
order $2$, so no linear combination is needed.
In the multivariate setting, the variance factor for a cross-term
$S_{x_s x_t}$ is the product
$\sigma_{1,1}^2 \cdot \sigma_{1,1}^2 = \tfrac{25}{16}$,
while the variance factor for a square term $S_{x_s^2}$ is $\sigma_{2,1}^2 \cdot 1 = 1$.
These are the factors for the active coordinates only;
the full tensor additionally contributes $3^{p-2}$ for a cross-term $S_{x_s x_t}$, and $3^{p-1}$ for a square term $S_{x_s^2}$ from the unused feature coordinates.
More generally, if $\tilde B_k=B_k+Z_k$ with independent
$Z_k\sim\mathrm{Lap}(\tau)$, then the recovered monomial sum
$\hat S_{x^m}=\sum_{k=m}^{2d}w_{m,k}\tilde B_k$ satisfies
\begin{align*}
  \mathrm{Var}(\hat S_{x^m}-S_{x^m})
  &= \sum_{k=m}^{2d}w_{m,k}^2\,\mathrm{Var}(Z_k)\\
  &=2\tau^2\sigma_{m,d}^2.
\end{align*}
Thus $\sigma_{m,d}^2$ is the variance amplification relative
to a single Laplace draw of scale $\tau$.

\paragraph{Recovering sufficient statistics.}
Since~\eqref{eq:release} is post-processed deterministically,
privacy is already guaranteed; we now show that unbiased estimates of all entries of
$X^\top X$ and $X^\top \bm y$ are recoverable as linear
functions of $(\tBcell{\bm k}{r})_{\bm k, r}$. In the following equations,
$\bm x=(x_1,\ldots,x_p)$ denotes the vector of $p$ features, and
$\bm m$ specifies their exponents. For OLS, $\bm x^{\bm m}$ is
needed only for $1$, $x_j$, $x_j^2$, and $x_jx_\ell$
($1\le j<\ell\le p$), so $\sum_j m_j\le2$.
For $\bm x^{\bm m'}y$, only $y$ and $x_jy$ are needed,
so $\sum_j m'_j\le1$. The general recovery formulas are:
\begin{align}
  \hat n
  &:= \sum_{\bm k,\, r} \tBcell{\bm k}{r},
  \label{eq:est_n}\\
  \hat S_{\bm x^{\bm m}}
  &:= \sum_{\bm k,\, r}
    \Bigl(\prod_{s=1}^p w_{m_s, k_s}\Bigr)\,
    \tBcell{\bm k}{r},
  \label{eq:est_xm}\\
  \hat S_{\bm x^{\bm m'} y}
  &:= \sum_{\bm k}
    \Bigl(\prod_{s=1}^p w_{m'_s, k_s}\Bigr)\,
    \tBcell{\bm k}{y}.
  \label{eq:est_xmy}\\
  \color{black}\hat S_y
  &\color{black}:= \sum_{\bm k}\tBcell{\bm k}{y}.
  \label{eq:est_y}
\end{align}
Each estimator is unbiased: $\mathbb{E}[\hat S_{\bm x^{\bm m}}]
= S_{\bm x^{\bm m}}$ and $\mathbb{E}[\hat S_{\bm x^{\bm m'}y}]
= S_{\bm x^{\bm m'}y}$, since the Laplace noise has mean zero and
the recovery weights satisfy the monomial identity
(Lemma~\ref{lem:bernstein}).
The variance of each estimation error follows directly from the
independence of the Laplace draws and the definition of the variance
factor $\sigma_{m,d}^2$:
\begingroup
\allowdisplaybreaks[4]
\begin{align}
  \mathrm{Var}\!\bigl(\hat S_{\bm x^{\bm m}} - S_{\bm x^{\bm m}}\bigr)
  &= \frac{4}{\varepsilon^2}
     \prod_{s=1}^{p} \sigma_{m_s,1}^2,
  \label{eq:var_xm}\\[4pt]
  \mathrm{Var}\!\bigl(\hat S_{\bm x^{\bm m'} y} - S_{\bm x^{\bm m'} y}\bigr)
  &= \frac{2}{\varepsilon^2}
     \prod_{s=1}^{p} \sigma_{m'_s,1}^2,
  \label{eq:var_xmy}\\[4pt]
  \mathrm{Var}\!\bigl(\hat n - n\bigr)
  &= \frac{4 \cdot 3^p}{\varepsilon^2},
  \label{eq:var_n}\\[4pt]
  \color{black}\mathrm{Var}(\hat S_y-S_y)
  &\color{black}=\frac{2\cdot3^p}{\varepsilon^2}.
  \label{eq:response_sum_explicit}
\end{align}
\endgroup
Feature-only statistics sum both response
cells ($y$ and $1-y$), contributing an additional factor of two, whereas response-weighted
statistics use only the $y$ cells.

\paragraph{Variance comparison with DP-RSS.}
Table~\ref{tab:PROBES_vs_dprss} compares the per-statistic estimation-error variance
of \methodname{} against DP-RSS~\citep{harini2026refined} for the
univariate case $p=1$.\footnote{DP-RSS~\citep{harini2026refined}
maps each record $(x,y)\in[0,1]^2$ to the simplex vectors
$(x,y)\mapsto [(x^2,x-x^2,1-x)$, $(xy,(1-x)y,1-y)]$. 
Each tuple has add/remove $\ell_1$-sensitivity one and receives budget
$\varepsilon/2$. Their noisy sums provide independent count estimates;
algebraic identities then yield independent estimates of each target
moment, combined by inverse-variance weighting without additional
privacy budget.}
 The variance with \methodname{} matches DP-RSS on $\hat n$
and strictly improves on every
other statistic, with gains ranging from $1.67\times$ on $\hat S_{x^2}$
to $2.67\times$ on $\hat S_{xy}$.

\begingroup\color{black}
\begin{table}[t]
\centering
\caption{Per-statistic estimation-error variance
$\mathrm{Var}(\hat{\cdot}-\cdot)$ at $p=1$, $d=1$:
\methodname{} vs.\ DP-RSS~\citep{harini2026refined}.}
\label{tab:PROBES_vs_dprss}
\begin{tabular}{l|c|c|c}
\toprule
Statistic & DP-RSS & \methodname{} & Improvement \\
\midrule
$\hat n$        & $12/\varepsilon^2$     & $12/\varepsilon^2$     & $1.00\times$ \\
$\hat S_{x^2}$  & $20/(3\varepsilon^2)$  & $4/\varepsilon^2$      & $1.67\times$ \\
$\hat S_{xy}$   & $20/(3\varepsilon^2)$  & $5/(2\varepsilon^2)$   & $2.67\times$ \\
$\hat S_{x}$    & $32/(3\varepsilon^2)$  & $5/\varepsilon^2$      & $2.13\times$ \\
$\hat S_{y}$    & $32/(3\varepsilon^2)$  & $6/\varepsilon^2$      & $1.78\times$ \\
\bottomrule
\end{tabular}
\end{table}\endgroup

\paragraph{Variance comparison with a joint Laplace release.}
For the complete OLS workload in~\eqref{eq:suff_stats}, including the
private count, there are $K=(p+1)(p+4)/2$ coordinates. Under add/remove
adjacency and fixed public preprocessing into $[0,1]^{p+1}$, the joint
$\ell_1$-sensitivity is $K$. A joint Laplace mechanism makes one release of this
vector at the full budget $\varepsilon$, with independent
$\Lap(K/\varepsilon)$ coordinates. Consequently, for every sufficient
statistic $S$ in this workload,
\begin{equation}
  \Var(\hat S_{\rm L}-S)
  =\frac{2K^2}{\varepsilon^2}
  =\frac{(p+1)^2(p+4)^2}{2\varepsilon^2},
  \label{eq:naive_joint_variance}
\end{equation}
where $\hat S_{\rm L}$ denotes the private estimate of statistics $S$ using joint Laplace release. On the other hand, equations~\eqref{eq:var_xm}--\eqref{eq:response_sum_explicit} give the full-tensor
PROBES variances at order $2$ ($d=1$). Thus the direct Laplace release has
$\Theta(p^4/\varepsilon^2)$ variance, whereas PROBES has
$\Theta(3^p/\varepsilon^2)$ variance. For a fixed dataset and sufficient
statistic $S$, define the ratio of estimation-error variances
\[
  \nu_p(S)=\frac{\Var(\hat S-S)}{\Var(\hat S_{\rm L}-S)}.
\]

\begin{table}[t]
\color{black}
\captionsetup{font={color=black},labelfont={color=black}}
\centering
\small
\setlength{\tabcolsep}{3pt}
\caption{$\nu_p(S)$: Ratio of estimation-error variances with PROBES and
a joint Laplace release for OLS statistic $S$ ($d=1$).}
\label{tab:probes_naive_ratios}
\begin{tabular}{lccccc}
\toprule
Statistic & \shortstack{PROBES\\$\varepsilon^2\Var(\hat S-S)$} & $\nu_{4}(S)$ & $\nu_{6}(S)$ & $\nu_{8}(S)$ & $\nu_{10}(S)$ \\
\midrule
$n$ & $4\cdot3^p$ & 0.405 & 1.190 & 4.500 & 19.919 \\
$S_{x_j}$ & $5\cdot3^{p-1}$ & 0.169 & 0.496 & 1.875 & 8.299 \\
$S_{x_j^2}$ & $4\cdot3^{p-1}$ & 0.135 & 0.397 & 1.500 & 6.640 \\
$S_{x_jx_\ell}$, $j\ne\ell$ & $\frac{25}{4}3^{p-2}$ & 0.070 & 0.207 & 0.781 & 3.458 \\
$S_y$ & $2\cdot3^p$ & 0.203 & 0.595 & 2.250 & 9.959 \\
$S_{x_jy}$ & $\frac{5}{2}3^{p-1}$ & 0.084 & 0.248 & 0.938 & 4.150 \\
\bottomrule
\end{tabular}
\end{table}

Table~\ref{tab:probes_naive_ratios} gives the per-statistic ratios
at $p=4,6,8,10$ under this same joint workload and full privacy budget.
The PROBES column gives $\varepsilon^2\Var(\hat S-S)$; the joint Laplace value
is $2K^2$, with $K=(p+1)(p+4)/2$ including the private count.
Ratios below $1$ favor PROBES, while ratios above $1$ favor the direct Laplace release;
they are independent of $\varepsilon$ and rounded to three
decimals. In every displayed row, $\nu_4(S)<1$ and $\nu_{10}(S)>1$:
PROBES has lower per-statistic variance at $p=4$, whereas the direct release
has lower per-statistic variance at $p=10$.

\paragraph{ Regularisation and Ada-\methodname{}}
Privacy noise can make
the recovered matrix unsuitable for stable inversion. Eigenvalues describe
how a symmetric matrix scales different directions. Unlike the original Gram
matrix, its noisy version can have negative eigenvalues; eigenvalues near zero
can also cause inversion to excessively amplify the corresponding directions.
To address these problems, we first add enough to the diagonal to make all
eigenvalues non-negative, then add a positive margin that keeps them away from
zero and limits this amplification. 
 We address the regularisation problem via the regulariser
\begin{equation}
  \widetilde{\bm\theta}
  \;:=\;
  \bigl(\hat A - \lambda_{\min}^{-}\,I + \rho\,I\bigr)^{-1}
  \hat{\bm b},
  \label{eq:ridge_solve}
\end{equation}
where $\lambda_{\min}^{-} := \min(\lambda_{\min}(\hat A), 0) \le 0$
is the most negative eigenvalue of $\hat A$ (zero if $\hat A$ is
already positive semidefinite), and
$\rho := (2/\varepsilon)\cdot 3^{(p-1)/2}$ is set proportional to
the standard deviation of the noise of $\hat S_{x^2}$.
Together they guarantee $\hat A - \lambda_{\min}^{-}I + \rho I
\succeq \rho I \succ 0$, making~\eqref{eq:ridge_solve} always
well-defined.  We refer to this variant as \emph{Ada-\methodname{}};
its full pseudocode is given in Algorithm~\ref{alg:rPROBES}.

\begingroup\color{black}
\paragraph{Open problem: optimal $\lambda$ under correlated noise.}
\label{sec:open_lambda}
A fundamental open question is how to choose $\lambda$ \emph{optimally}
when the noise added to the sufficient statistics is correlated across
entries. In \methodname{}, the Laplace perturbation is applied to a
single joint vector whose coordinates correspond to all entries
of~\eqref{eq:suff_stats} simultaneously. Although the Laplace noises added to individual Bernstein cells are independent, the recovered sufficient statistics are generally correlated because multiple statistics are formed from overlapping sets of noisy cells. Formally, one would seek
\begin{equation}
  \lambda^{*}
  \;=\;
  \operatorname*{arg\,min}_{\lambda > 0}\;
  \mathbb{E}\bigl[
    \|\hat{\bm\theta}_\lambda - \bm\theta^*\|_2^2
    \;\big|\;
    \Sigma_{\mathrm{noise}}
  \bigr],
  \label{eq:opt_lambda}
\end{equation}
where the expectation is over the Laplace noise realisation and
$\bm\theta^*$ is the true coefficient vector.
Existing heuristics (e.g., setting $\lambda$ proportional to the
diagonal noise variance) ignore the off-diagonal correlations and may
therefore be suboptimal. Developing a principled, data-adaptive, and
privacy-compatible selection rule for $\lambda$ that accounts for the
full noise covariance structure of jointly released sufficient statistics
remains an open problem of both theoretical and practical importance.
\endgroup

\begingroup\color{black}
\begin{algorithm}[t]
\caption{\methodname{}: Single-Shot Bernstein Simplex Release}
\label{alg:PROBES}
\SetKwInOut{Input}{Input}
\SetKwInOut{Privacy}{Privacy}
\SetKwInOut{Output}{Output}
\Input{Dataset $D = \{(\bm{x}_i,y_i)\}_{i=1}^n$; inputs in $[0,1]^{p+1}$.}
\Privacy{Budget $\varepsilon > 0$}
\Output{Privatised cell vector $\{\tBcell{\bm{k}}{r}\}$ of size $3^p \cdot 2$}
\BlankLine
$p_{k_s}(x) \gets \binom{2}{k_s} x^{k_s}(1-x)^{2-k_s}$ for $k_s \in \{0,1,2\}$\;
$c_y(y) \gets y$,\quad $c_{1-y}(y) \gets 1-y$\;
$\Bcell{\bm{k}}{r} \gets 0$ for all $\bm{k} \in \{0,1,2\}^p$, $r \in \{y, 1-y\}$\;
\For{each record $(\bm{x}_i, y_i) \in D$}{
  \For{each $\bm{k} \in \{0,1,2\}^p$ and $r \in \{y, 1-y\}$}{
    $\Bcell{\bm{k}}{r} \gets \Bcell{\bm{k}}{r} + \Bigl(\prod_{s=1}^{p} p_{k_s}(x_{is})\Bigr)\cdot c_r(y_i)$\;
  }
}
\For{each $(\bm{k}, r)$}{
  Draw $Z_{\bm{k},r} \overset{\mathrm{iid}}{\sim} \mathrm{Lap}(1/\varepsilon)$\;
  $\tBcell{\bm{k}}{r} \gets \Bcell{\bm{k}}{r} + Z_{\bm{k},r}$\;
}
\Return{$\{\tBcell{\bm{k}}{r}\}_{\bm{k} \in \{0,1,2\}^p,\, r \in \{y, 1-y\}}$}\;
\end{algorithm}\endgroup

\begingroup\color{black}
\begin{algorithm}[t]
\caption{Ada-\textsc{PROBES}: Post-Processing OLS Extraction ($d=1$)}
\label{alg:rPROBES}
\SetKwInOut{Input}{Input}
\SetKwInOut{Output}{Output}
\Input{PROBES release $\{\tBcell{\bm{k}}{r}\}$; dimension $p$; privacy budget $\varepsilon$}
\Output{Private OLS coefficients $\widetilde{\bm{\theta}}$}
\BlankLine
$(w_{0,\cdot}, w_{1,\cdot}, w_{2,\cdot}) \gets (1,1,1),\ (0,\tfrac{1}{2},1),\ (0,0,1)$\;
$\hat{n} \gets \sum_{\bm{k},\, r} \tBcell{\bm{k}}{r}$\;
\For{each multi-index $\bm{m} \in \{0,1,2\}^p$}{
  $\hat{S}_{\bm{x}^{\bm{m}}} \gets \sum_{\bm{k},\,r}\Bigl(\prod_{s=1}^{p} w_{m_s,k_s}\Bigr)\tBcell{\bm{k}}{r}$\;
  $\hat{S}_{\bm{x}^{\bm{m}}y} \gets \sum_{\bm{k}}\Bigl(\prod_{s=1}^{p} w_{m_s,k_s}\Bigr)\tBcell{\bm{k}}{y}$\;
}
$\hat A_{00} \gets \hat n$\;
\For{$j = 1$ \KwTo $p$}{
  $\hat A_{0j} \gets \hat A_{j0} \gets \hat S_{x_j}$;\quad
  $\hat A_{jj} \gets \hat S_{x_j^2}$\;
}
\For{$j = 1$ \KwTo $p-1$, $k = j+1$ \KwTo $p$}{
  $\hat A_{jk} \gets \hat A_{kj} \gets \hat S_{x_jx_k}$\;
}
$\hat b_0 \gets \hat S_y$\;
\For{$j = 1$ \KwTo $p$}{
  $\hat b_j \gets \hat S_{x_jy}$\;
}
$\rho \gets (2/\varepsilon)\cdot 3^{(p-1)/2}$\;
$\lambda_{\min}^{-} \gets \min\bigl(\lambda_{\min}(\hat{A}),\, 0\bigr)$\;
$\widetilde{\bm{\theta}} \gets \bigl(\hat{A} - \lambda_{\min}^{-} I + \rho I\bigr)^{-1} \hat{\bm{b}}$\;
\Return{$\widetilde{\bm{\theta}}$}\;
\end{algorithm}\endgroup

\begingroup\color{black}
\begin{table}[t]
\centering
\caption{Real-world datasets: dataset identifiers, raw feature dimension $d$,
  train/test sizes after an 80/20 split, and prediction target.}
\label{tab:datasets}
\footnotesize
\setlength{\tabcolsep}{2.5pt}
\renewcommand{\arraystretch}{1.10}
\begin{tabularx}{\columnwidth}{@{}l >{\raggedright\arraybackslash}X c r r >{\raggedright\arraybackslash}X@{}}
\toprule
\textbf{ID} & \textbf{Dataset} & $d$ & $n_{\mathrm{train}}$ & $n_{\mathrm{test}}$ & \textbf{Target} \\
\midrule
D1 & California Housing~\citep{pace1997sparse}     &  8 & 16{,}512 &  4{,}128 & Median house value \\
D2 & NYC Yellow Taxi~\citep{nyctlc2024}            &  8 &  1{,}000 &    250   & Total fare amount \\
D3 & Appliances Energy~\citep{candanedo2017data}   & 26 & 15{,}788 &  3{,}947 & Appliances (Wh) \\
D4 & Super\-conductivity~\citep{hamidieh2018data}    & 81 & 17{,}010 &  4{,}253 & Critical temp.\ $T_c$ (K) \\
D5 & Year Prediction MSD~\citep{bertin2011million} & 90 & 50{,}000 & 12{,}500 & Song release year \\
D6 & Covertype~\citep{blackard1999comparative} & 54 & 100{,}000 & 25{,}000 & Elevation (m) \\
\bottomrule
\end{tabularx}
\end{table}\endgroup

\begin{table*}[!t]
\begingroup\color{black}
\fontsize{8}{9}\selectfont
\setlength{\tabcolsep}{2pt}
\renewcommand{\arraystretch}{0.95}
\centering
\caption{Real-world test RMSE (D1--D5; $\varepsilon=0.5,1,10$).}
\label{tab:realworld_combined}
\footnotesize
\setlength{\tabcolsep}{1.5pt}
\begin{tabular*}{\textwidth}{@{\extracolsep{\fill}}lrr|rrrrr|rrrrr|rrrrr@{}}
\toprule
 & & & \multicolumn{5}{c|}{$\varepsilon=0.5$} & \multicolumn{5}{c|}{$\varepsilon=1$} & \multicolumn{5}{c}{$\varepsilon=10$} \\
\cmidrule(lr){4-8}\cmidrule(lr){9-13}\cmidrule(lr){14-18}
ID & $p$ & \textsc{OLS} & \textsc{Ada-P} & \textsc{AdaSSP} & \textsc{DPrivLib} & \textsc{BinAgg} & \textsc{Chen} & \textsc{Ada-P} & \textsc{AdaSSP} & \textsc{DPrivLib} & \textsc{BinAgg} & \textsc{Chen} & \textsc{Ada-P} & \textsc{AdaSSP} & \textsc{DPrivLib} & \textsc{BinAgg} & \textsc{Chen} \\
\midrule
D1 & 1 & 0.842 & \textbf{0.842} & 0.879 & \underline{0.843} & 0.850 & 1.022 & \textbf{0.842} & 0.848 & \textbf{0.842} & \underline{0.844} & 0.937 & \textbf{0.842} & \textbf{0.842} & \textbf{0.842} & \textbf{0.842} & \underline{0.849} \\
D1 & 2 & 0.838 & \textbf{0.841} & 0.933 & 972.960 & \underline{0.867} & 1.065 & \textbf{0.839} & 0.884 & 366.450 & \underline{0.854} & 0.970 & \textbf{0.838} & 0.841 & 0.842 & \underline{0.839} & 0.850 \\
D1 & 3 & 0.833 & \textbf{0.838} & 0.963 & 2668.800 & \underline{0.866} & 1.111 & \textbf{0.836} & 0.905 & 750.700 & \underline{0.850} & 1.009 & \textbf{0.833} & 0.838 & 4.893 & \underline{0.834} & 0.852 \\
D1 & 4 & 0.806 & \textbf{0.811} & 0.982 & 4724.900 & \underline{0.852} & 1.137 & \textbf{0.809} & 0.915 & 1065.100 & \underline{0.825} & 1.031 & \textbf{0.806} & 0.812 & 3.795 & \underline{0.808} & 0.835 \\
D1 & 5 & 0.799 & \textbf{0.828} & 1.006 & $1.77{\times}10^{4}$ & \underline{0.863} & 1.163 & \textbf{0.819} & 0.941 & 5774.900 & \underline{0.835} & 1.052 & \textbf{0.803} & 0.815 & 113.730 & \underline{0.809} & 0.841 \\
D1 & 6 & 0.746 & \textbf{0.807} & 1.032 & $3.19{\times}10^{4}$ & \underline{0.886} & 1.186 & \textbf{0.776} & 0.968 & $1.43{\times}10^{4}$ & \underline{0.832} & 1.075 & \textbf{0.746} & 0.779 & 320.400 & \underline{0.757} & 0.813 \\
\midrule
D2 & 1 & 3.193 & \textbf{5.233} & \underline{14.064} & 18.145 & 17.542 & 16.405 & \textbf{4.103} & 11.618 & 12.085 & \underline{9.600} & 14.849 & \textbf{3.221} & 5.993 & \underline{3.490} & 3.594 & 11.677 \\
D2 & 2 & 2.914 & \textbf{5.446} & \underline{13.441} & $1.02{\times}10^{5}$ & 132.100 & 16.734 & \textbf{4.238} & \underline{11.610} & $3.60{\times}10^{4}$ & 19.819 & 15.244 & \textbf{3.005} & 5.921 & 1144.900 & \underline{4.281} & 11.474 \\
D2 & 3 & 2.402 & \textbf{10.672} & \underline{14.376} & $3.25{\times}10^{5}$ & 75.939 & 17.014 & \textbf{6.959} & \underline{12.255} & $1.47{\times}10^{5}$ & 27.889 & 15.619 & \textbf{2.835} & 6.575 & 4517.400 & \underline{5.373} & 11.815 \\
D2 & 4 & 0.824 & \textbf{13.899} & \underline{14.205} & $6.69{\times}10^{5}$ & 139.350 & 17.446 & \textbf{9.534} & \underline{12.196} & $2.95{\times}10^{5}$ & 46.521 & 16.056 & \textbf{2.407} & 6.800 & $1.17{\times}10^{4}$ & \underline{5.949} & 12.047 \\
D2 & 5 & 0.806 & 20.718 & \textbf{14.324} & $1.29{\times}10^{6}$ & 117.580 & \underline{16.950} & \underline{15.273} & \textbf{12.735} & $6.27{\times}10^{5}$ & 32.477 & 15.598 & \textbf{3.718} & 7.704 & $3.01{\times}10^{4}$ & \underline{6.878} & 12.148 \\
D2 & 6 & 0.805 & 29.055 & \textbf{14.370} & $2.46{\times}10^{6}$ & 179.620 & \underline{17.530} & 20.653 & \textbf{12.943} & $1.05{\times}10^{6}$ & 132.370 & \underline{16.111} & \textbf{4.945} & 8.593 & $5.25{\times}10^{4}$ & \underline{6.485} & 12.415 \\
\midrule
D3 & 1 & 97.553 & \textbf{97.573} & 99.116 & \underline{97.884} & 102.508 & 99.930 & \textbf{97.560} & 98.323 & \underline{97.725} & 98.531 & 98.802 & \textbf{97.553} & 97.558 & \underline{97.554} & 97.560 & 97.614 \\
D3 & 2 & 95.995 & \textbf{96.040} & 99.401 & \underline{97.475} & 101.279 & 100.469 & \textbf{96.011} & 97.811 & \underline{96.365} & 98.928 & 98.663 & \textbf{95.996} & \underline{96.017} & \textbf{95.996} & 96.041 & 96.242 \\
D3 & 3 & 95.941 & \textbf{96.193} & \underline{99.798} & 101.033 & 102.806 & 100.840 & \textbf{96.019} & 98.174 & \underline{97.455} & 98.936 & 98.994 & \textbf{95.945} & 95.982 & \underline{95.949} & 96.021 & 96.306 \\
D3 & 4 & 95.936 & \textbf{96.654} & \underline{99.622} & $8.75{\times}10^{4}$ & 106.358 & 101.166 & \textbf{96.150} & \underline{98.647} & $5.88{\times}10^{4}$ & 101.263 & 99.343 & \textbf{95.942} & 96.041 & \underline{95.988} & 96.086 & 96.453 \\
D3 & 5 & 95.663 & \textbf{98.444} & \underline{100.267} & $2.22{\times}10^{6}$ & 106.987 & 101.230 & \textbf{96.710} & \underline{99.003} & $4.86{\times}10^{5}$ & 101.396 & 99.472 & \textbf{95.680} & 96.210 & 96.068 & \underline{95.969} & 96.431 \\
D3 & 6 & 95.637 & 101.980 & \textbf{99.934} & $5.84{\times}10^{6}$ & 109.420 & \underline{101.862} & \textbf{97.837} & \underline{98.867} & $1.05{\times}10^{6}$ & 102.155 & 99.840 & \textbf{95.686} & 96.204 & 96.776 & \underline{96.053} & 96.499 \\
\midrule
D4 & 1 & 23.338 & \textbf{23.338} & 23.415 & \underline{23.342} & 23.559 & 25.221 & \textbf{23.338} & 23.360 & \underline{23.339} & 23.411 & 24.018 & \textbf{23.338} & 23.395 & \textbf{23.338} & \underline{23.341} & 23.352 \\
D4 & 2 & 23.311 & \textbf{23.319} & 23.897 & \underline{23.620} & 24.435 & 25.236 & \textbf{23.313} & 23.660 & \underline{23.385} & 23.865 & 24.168 & \textbf{23.311} & 23.319 & \underline{23.312} & 23.333 & 23.397 \\
D4 & 3 & 22.505 & \textbf{22.645} & \underline{24.078} & $4.43{\times}10^{4}$ & 24.539 & 25.504 & \textbf{22.559} & \underline{23.747} & 2123.400 & 23.914 & 24.362 & \textbf{22.507} & 22.784 & \underline{22.516} & 22.691 & 22.965 \\
D4 & 4 & 21.890 & \textbf{22.086} & \underline{23.373} & $8.37{\times}10^{4}$ & 24.125 & 25.143 & \textbf{21.964} & \underline{22.958} & $3.43{\times}10^{4}$ & 22.989 & 23.709 & \textbf{21.890} & 22.111 & \underline{21.910} & 21.972 & 22.251 \\
D4 & 5 & 21.746 & \textbf{22.171} & \underline{23.268} & $2.36{\times}10^{5}$ & 23.862 & 25.294 & \textbf{21.921} & \underline{22.863} & $6.42{\times}10^{4}$ & 23.029 & 23.682 & \textbf{21.748} & 22.016 & \underline{21.791} & 21.830 & 22.126 \\
D4 & 6 & 21.688 & \textbf{22.321} & \underline{23.237} & $5.68{\times}10^{5}$ & 24.178 & 25.389 & \textbf{21.973} & \underline{22.724} & $1.28{\times}10^{5}$ & 23.055 & 23.641 & \textbf{21.696} & 21.938 & 426.440 & \underline{21.759} & 22.028 \\
\midrule
D5 & 1 & 10.507 & \textbf{10.507} & 10.553 & \underline{10.510} & 10.613 & 10.999 & \textbf{10.507} & 10.521 & \underline{10.508} & 10.561 & 10.685 & \textbf{10.507} & \textbf{10.507} & \textbf{10.507} & \underline{10.514} & 10.528 \\
D5 & 2 & 10.426 & \textbf{10.426} & 10.644 & \underline{10.439} & 10.717 & 11.397 & \textbf{10.426} & 10.526 & \underline{10.431} & 10.604 & 10.823 & \textbf{10.426} & \underline{10.427} & \textbf{10.426} & 10.446 & 10.467 \\
D5 & 3 & 10.230 & \textbf{10.235} & 10.771 & \underline{10.336} & 11.247 & 11.776 & \textbf{10.232} & 10.576 & \underline{10.253} & 10.817 & 10.958 & \textbf{10.230} & 10.234 & \underline{10.231} & 10.274 & 10.305 \\
D5 & 4 & 10.231 & \textbf{10.252} & \underline{10.810} & 1098.400 & 11.229 & 12.049 & \textbf{10.236} & 10.617 & \underline{10.332} & 10.872 & 11.109 & \textbf{10.230} & 10.245 & \underline{10.232} & 10.322 & 10.327 \\
D5 & 5 & 10.210 & \textbf{10.303} & \underline{10.847} & $9.89{\times}10^{4}$ & 11.343 & 12.262 & \textbf{10.245} & \underline{10.699} & $1.74{\times}10^{4}$ & 10.946 & 11.222 & \textbf{10.210} & 10.293 & \underline{10.217} & 10.399 & 10.339 \\
D5 & 6 & 10.206 & \textbf{10.369} & \underline{10.777} & $2.26{\times}10^{5}$ & 11.261 & 12.306 & \textbf{10.276} & \underline{10.654} & $5.89{\times}10^{4}$ & 10.950 & 11.230 & \textbf{10.206} & 10.297 & \underline{10.216} & 10.360 & 10.336 \\
\bottomrule
\end{tabular*}

\endgroup
\par\medskip
\begingroup\color{black}
\centering
\caption{Test MSE on i.i.d.\ synthetic data at $\varepsilon = 1.0$.
  All entries are in units of $10^{-3}$.}
\label{tab:synthetic_iid_combined}
\footnotesize
\renewcommand{\arraystretch}{1.05}
\setlength{\tabcolsep}{1.5pt}

\begin{tabular*}{\textwidth}{@{\extracolsep{\fill}}c|rrrrr|rrrrr|rrrrr@{}}
\toprule
& \multicolumn{5}{c|}{$n = 2{,}000p$}
  & \multicolumn{5}{c|}{$n = 5{,}000p$}
  & \multicolumn{5}{c}{$n = 10{,}000p$} \\
\cmidrule(lr){2-6}\cmidrule(lr){7-11}\cmidrule(lr){12-16}
$p$
  & \textsc{Ada-P} & \textsc{AdaSSP} & \textsc{DPrivLib} & \textsc{BinAgg} & \textsc{Chen}
  & \textsc{Ada-P} & \textsc{AdaSSP} & \textsc{DPrivLib} & \textsc{BinAgg} & \textsc{Chen}
  & \textsc{Ada-P} & \textsc{AdaSSP} & \textsc{DPrivLib} & \textsc{BinAgg} & \textsc{Chen} \\
\midrule
1
  & \best{0.911} & 4.210 & \second{1.000} & 1.960 & 52.553
  & \best{0.897} & 1.350 & \second{0.913} & 1.280 & 20.587
  & \best{0.906} & 1.030 & \second{0.910} & 1.080 & 9.051 \\
2
  & \best{0.902} & 2.540 & \second{1.020} & 1.840 & 27.709
  & \best{0.907} & 1.360 & \second{0.922} & 1.190 & 9.707
  & \best{0.900} & 0.987 & \second{0.904} & 1.020 & 4.219 \\
3
  & \best{0.909} & 1.880 & \second{1.070} & 1.840 & 21.002
  & \best{0.905} & 1.240 & \second{0.925} & 1.170 & 7.234
  & \best{0.899} & 0.991 & \second{0.905} & 0.999 & 3.207 \\
4
  & \best{0.919} & 1.770 & \second{1.120} & 1.590 & 17.799
  & \best{0.900} & 1.210 & \second{0.930} & 1.220 & 6.136
  & \best{0.901} & 0.989 & \second{0.910} & 1.050 & 2.792 \\
5
  & \best{0.946} & 1.680 & \second{1.250} & 1.890 & 15.980
  & \best{0.905} & 1.180 & \second{0.957} & 1.210 & 5.571
  & \best{0.905} & 0.996 & \second{0.918} & 1.040 & 2.595 \\
6
  & \best{1.000} & 1.610 & \second{1.330} & 2.100 & 14.930
  & \best{0.921} & 1.130 & \second{0.977} & 1.180 & 5.276
  & \best{0.905} & 1.000 & \second{0.918} & 1.050 & 2.482 \\
\bottomrule
\end{tabular*}
\endgroup
\end{table*}

\section{Experimental Evaluation}
\label{sec:experiments}

We evaluate \textsc{Ada-PROBES} on
six real-world regression benchmarks and a controlled synthetic suite,
comparing against \textsc{AdaSSP}~\cite{wang2018PrivateLinearRegression},
\textsc{BinAgg}~\cite{lin2025dplrsynthetic}, and
\textsc{DiffPrivLib}~\cite{holohan2019diffprivlib} (corrected
per~\cite{cebere2026bug}).
We also compare with Chen et al.~\citep{chen2025beyond}, and run
Algorithms 4 and 5 in~\citep{chen2025beyond} by setting constant $C=4$. We treat this multiplier $C$ as a hyperparameter selected by
comparing test RMSE, so these are exploratory, test-tuned comparisons.
This tuning requirement is a practical limitation of the algorithm:
stable matrix inversion must be balanced against prediction accuracy.
Test RMSE (real-world datasets) and test MSE on a held-out 20\% split
(synthetic, 100 Monte Carlo
seeds) are reported on the original scale.
\textsc{AdaSSP} satisfies $(\varepsilon,\delta)$-DP with
$\delta = \min(10^{-6},\,n_{\mathrm{train}}^{-2})$;
\textsc{BinAgg} uses Gaussian-DP parameterised by $\mu$, converted from
$(\varepsilon,\delta)$-DP.
We report the comparison results below.\footnote{%
Bold and underlined entries in numerical comparison tables identify the lowest and second-lowest distinct 
RMSE values among available private methods in each row and privacy-budget group; OLS is the non-private baseline.
}

\subsection{Real-World Datasets}
\label{ssec:real_datasets}

  \paragraph{Methods.}
  \textsc{PROBES} releases all Bernstein cells in one Laplace vector draw at budget $\varepsilon$.
  \textsc{Ada-PROBES} is the practical variant used in all experiments: it runs Algorithm~\ref{alg:PROBES} and then Algorithm~\ref{alg:rPROBES}, i.e., the same release with the noise-calibrated regularised solve of~\eqref{eq:ridge_solve}.
  Competitors are \textsc{AdaSSP} ($(\varepsilon,\delta)$-DP), \textsc{BinAgg} (Gaussian-DP), and \textsc{DiffPrivLib} (Laplace noise based objective perturbation).

\paragraph{Datasets.}
Table~\ref{tab:datasets} summarises the six benchmarks.\footnote{For large datasets (D2, D5, D6), we subsample so that training
reaches a fixed target size after an 80/20 train-test split.}
Covertype (D6) is originally a classification dataset; we use its
continuous elevation attribute as a constructed regression target and
treat this setting as a supplementary evaluation.
Later, we use Covertype alongside a correlated synthetic benchmark to
evaluate S-PROBES, the polynomial-size extension of PROBES, at larger feature
dimensions (Section~\ref{sec:steiner_tmlr_main}).

\paragraph{Protocol and preprocessing.}
  For each dataset and dimensionality $p$, we select the top-$p$ features by absolute Pearson correlation using the training data only. This feature-selection procedure is used solely to construct controlled low-dimensional experiments; it is treated as external preprocessing and no privacy budget is therefore allocated towards this data-dependent task. For the experiments, we choose $p \in \{1,\ldots,6\}$ for D1--D6.

All methods operate under the \textbf{unit-cube protocol} i.e
$[0,1]$ normalisation, with coefficients mapped back to original scale
for evaluation. When data lie in
$[\mathbf{x}_{\min}, \mathbf{x}_{\max}] \times [y_{\min}, y_{\max}]$,
define
$x'_j = (x_j - x_{j,\min})/\Delta_{x_j}$ and
$y' = (y - y_{\min})/\Delta_y$.
Run \textsc{PROBES} on the normalised dataset to obtain private
coefficients $(\hat{\boldsymbol{\alpha}}', \hat{\beta}')$, then invert:
\[
  \hat{\alpha}_j
    = \frac{\Delta_y}{\Delta_{x_j}}\,\hat{\alpha}'_j,
  \qquad
  \hat{\beta}
    = y_{\min}
      + \Delta_y\!\left(
          \hat{\beta}' - \sum_j \frac{x_{j,\min}}{\Delta_{x_j}}\hat{\alpha}'_j
        \right).
\]

 For reproducibility and to reduce the number of features used in the
experiments, we perform feature selection on both synthetic and
real-world datasets using the Pearson correlation between each feature
and the response, computed using the training data only. The selected
feature subset is then applied unchanged to the corresponding test data,
followed by train-only min--max normalisation, with the resulting
transformation likewise reused for the test data.

We assume knowledge of the required dataset bounds for the normalisation procedure; and for fairness and consistency, the above preprocessing steps are applied across all the different private mechanisms compared in our work. Therefore, the privacy guarantee reported in this section, for different mechanisms, does not include the data-dependent feature selection and associated preprocessing.

Table~\ref{tab:realworld_combined} reports D1--D5 at
$\varepsilon=0.5,1,10$. The Covertype results are
reported in Section~\ref{sec:steiner_tmlr_main}.

\subsection{Synthetic Datasets}
\label{ssec:synthetic}

We evaluate all methods on   one synthetic benchmark where  every method operates on data normalised to
$[0,1]^{p+1}$, trains on 80\% of records, and is evaluated on the
held-out 20\% test split.
Results are mean MSE over 100 seeds at
$\varepsilon = 1.0$ .

\paragraph{IID benchmark.}
Features are drawn i.i.d.\ $\mathbf{x}_i \sim \mathrm{Uniform}(0,1)^p$
with response
$y_i = \mathrm{clip}\bigl(0.2 + \tfrac{0.6}{p}\sum_j x_{ij} +
\mathcal{N}(0,0.03^2),\,0,\,1\bigr)$.
Table~\ref{tab:synthetic_iid_combined} reports mean test MSE on the held-out 20\% split
for
  $p \in \{1,\ldots,6\}$  and $n \in \{2000p,\,5000p,\,10000p\}$
(100 Monte Carlo seeds).

 \subsection{Discussion}
\label{ssec:discussion}

  Across the evaluated datasets, \textsc{Ada-PROBES} has the lowest test error in most settings. Its advantage is strongest for moderate feature dimension and becomes smaller on the smallest dataset and at the largest evaluated $p$.

\paragraph{Performance across datasets.}

On three of six real-world datasets (D1, D4, D5),
\textsc{Ada-PROBES} achieves the lowest test RMSE across all evaluated
privacy budgets and feature dimensions. This includes the low $\varepsilon$ settings ($\varepsilon = 0.5$, $\varepsilon = 1$) where privacy constraints are most stringent. For instance, on D1 with $\varepsilon = 0.5$  and   $p = 4$, \textsc{Ada-PROBES} achieves RMSE $0.811$, within $0.6\%$ of the non-private OLS baseline, while \textsc{AdaSSP} reaches $0.982$ (a $21\%$ gap) and \textsc{DiffPrivLib} collapses to $4724.9$ .

The synthetic i.i.d.\ benchmark (Table~\ref{tab:synthetic_iid_combined}) reinforces this pattern: \textsc{Ada-PROBES} is best-performing across all feature dimensions $p \in \{1,\ldots,6\}$ and sample-size regimes, with improvements over \textsc{AdaSSP} ranging from $1.8\times$ to $2.5\times$ at low sample sizes.

\paragraph{  Strong advantage on small datasets with low dimensionality. }
\begingroup\color{black}
\begin{algorithm}[!t]
\color{black}
\SetAlCapFnt{\color{black}\small}
\SetAlCapNameFnt{\color{black}\small\bfseries}
\SetAlFnt{\fontsize{8.5}{10}\selectfont}
\SetInd{0.5em}{0.7em}
\SetAlgoHangIndent{1em}
\SetNlSty{red}{}{}

\caption{S-\textsc{Probes}-3: Full Pipeline \mbox{($p=6$, Fano-Plane Blocks)}}
\label{alg:sprobes_full}
\SetKwInOut{Input}{Input}\SetKwInOut{Output}{Output}
\Input{$\{(\mathbf{x}_i,y_i)\}_{i=1}^n\subset[0,1]^7$; $M=10$; $\eps>0$}
\Output{Private OLS vector $\widetilde{\boldsymbol\theta}\in\R^7$}
\BlankLine
\For{$\ell=1$ \KwTo $7$}{
  Let $B_\ell=\{a,b,c\}$; initialise $\Phi^{(B_\ell)}_{uvw}\gets0$, $(u,v,w)\in\{0,1\}^3$\;
  \For{$i=1$ \KwTo $n$}{
    $\Phi^{(B_\ell)}_{uvw}\mathrel{+}=a_i^u(1-a_i)^{1-u}$\newline
    $\qquad {}\cdot b_i^v(1-b_i)^{1-v}c_i^w(1-c_i)^{1-w}$\;
  }
  \For{$(u,v,w)\in\{0,1\}^3$}{
    $\tilde\Phi^{(B_\ell)}_{uvw}\gets\Phi^{(B_\ell)}_{uvw}+Z$,\quad $Z\overset{\mathrm{iid}}{\sim}\Lap(M/\eps)$\;
  }
  \textbf{release} $\tilde\Phi^{(B_\ell)}$\;
}
\For{$k=1$ \KwTo $3$}{
  Let $D_k=\{x_{2k-1},x_{2k}\}$; initialise $\Phi^{(D_k)}_{uv}\gets0$\;
  \For{$i=1$ \KwTo $n$}{
    $\Phi^{(D_k)}_{uv}\mathrel{+}=(x_{i,2k-1}^2)^u(1-x_{i,2k-1}^2)^{1-u}$\newline
    $\qquad {}\cdot(x_{i,2k}^2)^v(1-x_{i,2k}^2)^{1-v}$\;
  }
  \For{$(u,v)\in\{0,1\}^2$}{
    $\tilde\Phi^{(D_k)}_{uv}\gets\Phi^{(D_k)}_{uv}+Z$,\quad $Z\overset{\mathrm{iid}}{\sim}\Lap(M/\eps)$\;
  }
  \textbf{release} $\tilde\Phi^{(D_k)}$\;
}
\BlankLine
\For{each pair $(x_j,x_k)$, $j<k$}{
  $B_\ell\gets$ Steiner block containing $(x_j,x_k)$;\quad
  $\hat S_{x_jx_k}\gets\sum_{(u,v,w):\,u_{s_j}=1,u_{s_k}=1}\tilde\Phi^{(B_\ell)}_{uvw}$\;
}
\For{$j=1$ \KwTo $6$}{
  $B_\ell\gets$ block containing $(x_j,y)$;\quad
  $\hat S_{x_jy}\gets\sum_{(u,v,w):\,u_{s_j}=1,u_{s_y}=1}\tilde\Phi^{(B_\ell)}_{uvw}$\;
}
\For{$k=1$ \KwTo $3$}{
  $\hat S_{x_{2k-1}^2}\gets\tilde\Phi^{(D_k)}_{10}+\tilde\Phi^{(D_k)}_{11}$;\quad
  $\hat S_{x_{2k}^2}\gets\tilde\Phi^{(D_k)}_{01}+\tilde\Phi^{(D_k)}_{11}$\;
}
\For{each $v\in\{x_1,\ldots,x_6,y\}$ with containing blocks $B_{\ell_1},B_{\ell_2},B_{\ell_3}$}{
  $\hat S_v\gets(1/3)\sum_{t=1}^3\sum_{(u,v,w):\,u_{s_v}=1}\tilde\Phi^{(B_{\ell_t})}_{uvw}$\;
}
$\hat n\gets(1/3)\sum_{k=1}^3\!\bigl(\tilde\Phi^{(D_k)}_{00}+\tilde\Phi^{(D_k)}_{01}+\tilde\Phi^{(D_k)}_{10}+\tilde\Phi^{(D_k)}_{11}\bigr)$\;
\BlankLine
$\hat A_{00}\gets\hat n$;\quad
$\hat A_{jj}\gets\hat S_{x_j^2}$ ($j\geq1$)\;
$\hat A_{0j}=\hat A_{j0}\gets\hat S_{x_j}$;\quad
$\hat A_{jk}=\hat A_{kj}\gets\hat S_{x_jx_k}$ ($j<k$)\;
$\hat b_0\gets\hat S_y$;\quad $\hat b_j\gets\hat S_{x_jy}$ ($j=1,\ldots,6$)\;
\BlankLine
$\rho\gets\sqrt{2}\,M/\eps$;\quad $\lambda\gets\min(\lambda_{\min}(\hat A),0)$\;
$\widetilde{\boldsymbol\theta}\gets(\hat A-\lambda I+\rho I)^{-1}\hat{\mathbf{b}}$\;
\Return{$\widetilde{\boldsymbol\theta}$}\;
\end{algorithm}\endgroup

 On D2  (NYC Taxi, $n_{\text{train}} = 1000$), which represents a realistic small-sample regression scenario, \textsc{Ada-PROBES} delivers exceptional improvements at low dimensions. At $p=1$ and $\varepsilon=0.5$, \textsc{Ada-PROBES} achieves RMSE $5.233$, compared to \textsc{AdaSSP}'s $14.064$---a $2.7\times$ improvement. At $p=1$, $\varepsilon=1$, the advantage persists: \textsc{Ada-PROBES} $4.103$ versus \textsc{AdaSSP} $11.618$  (  $2.8\times$ better). Even at $p=2$ and tight privacy ($\varepsilon=0.5$), \textsc{Ada-PROBES} $5.446$ outperforms  \textsc{AdaSSP}   $13.441$ by a factor of $2.5\times$, and at $p=2$, $\varepsilon=1$, the gap widens to $2.7\times$ (Ada-PROBES $4.238$ vs. \   \textsc{AdaSSP} $11.610$). These are substantial practical improvements: on small-sample budgets with relatively small $p$, \textsc{Ada-PROBES} can cut estimation error nearly in half compared to competing methods. This advantage is especially valuable in real-world scenarios, such as rare disease datasets, where data samples are scarce and privacy constraints are stringent.

\paragraph{D2 at relatively higher feature dimension.}

On D2  (  NYC Taxi, $n_{\text{train}}=1000$), \textsc{Ada-PROBES} is best for
$p\leq4$; however, at $p=5$ and $\varepsilon=0.5$ ,   \textsc{AdaSSP}
achieves RMSE $14.324$ versus $20.718$ for \textsc{Ada-PROBES}. When datasets are small (order $O(10^3)$) and dimensions approach the Bernstein scaling limit ($p \ge 5$), the concentration of noise across $3^p \times 2$ cells in \textsc{Ada-PROBES} can outweigh the structural advantage of the unsplit budget.

\paragraph{Exception 2: D3 ($p=6$).}

On D3 at $p = 6$ and $\varepsilon = 0.5$, \textsc{AdaSSP} achieves RMSE $99.934$ (best) versus \textsc{Ada-PROBES} $101.980$ (second-best)---a narrow margin of $2\%$. At higher privacy budgets ($\varepsilon \geq 1$), \textsc{Ada-PROBES} recovers its advantage. This suggests that at the boundary of practical dimensionality ($p = 6$ corresponds to $3^6 \times 2 = 1458$ cells), and under very tight privacy ($\varepsilon = 0.5$), the advantage of the Bernstein representation can narrow when other methods have better eigenvalue control.

\paragraph{Behavior of \textsc{DiffPrivLib}.}

\textsc{DiffPrivLib} exhibits substantial numerical instability in many
of the higher-dimensional, low-privacy settings, with RMSE values reaching
the $10^3$--$10^6$ range.

\paragraph{Practical guidelines.}

\begin{itemize}[leftmargin=*, nosep, topsep=2pt]

\item {Use \textsc{Ada-PROBES} as the default:}
On large and moderate-sized datasets with $p \le 6$, \textsc{Ada-PROBES} consistently achieves low error. On small-sample datasets with low dimensionality ($p \le 3$), the method delivers particularly strong improvements---e.g., $2.7\times$ better than competing methods on D2 at $p=1$, $\varepsilon=0.5$---making it a strong option for the low-dimensional, privacy-constrained settings evaluated here.

\item {Monitor for small-sample, $p\ge 5$ crossover:}
When $n_{\text{train}} = O(10^3)$ and $p \ge 5$, consider \textsc{AdaSSP} as a secondary option. On the larger real-world datasets evaluated here, \textsc{Ada-PROBES} remains competitive across the tested privacy budgets.

  \end{itemize}

  These are conditional mechanism comparisons as the training-data feature
selection and empirical sample normalization is not charged to the reported privacy budget $\varepsilon$. End-to-end use either requires publicly available features and sample bounds, or a separately composed privacy budget.

 Overall, the experiments support a clear advantage for
\textsc{Ada-PROBES}, especially in the high-privacy regime with $p\le 4$. As the
number of features increases, the exponential growth of the Bernstein
representation progressively reduces this advantage. At the same
time, increasing the sample size reduces the relative impact of
privacy noise. Consequently, there is no single method that dominates
across all settings: the most effective mechanism depends jointly on
$p$, $\varepsilon$, sample size, and dataset characteristics.

\FloatBarrier
\begingroup\color{black}
\begin{table*}[!t]
\color{black}
\centering
\captionsetup{font={color=black},labelfont={color=black}}
\caption{Covertype test RMSE}
\label{tab:covertype_low}
\footnotesize
\setlength{\tabcolsep}{1.5pt}
\begin{tabular*}{\textwidth}{@{\extracolsep{\fill}}rr|rrrrrrrr|rrrrrrrr@{}}
\toprule
 & & \multicolumn{8}{c|}{$\varepsilon=1$} & \multicolumn{8}{c}{$\varepsilon=10$} \\
\cmidrule(lr){3-10}\cmidrule(lr){11-18}
$p$ & \textsc{OLS} & \textsc{Ada-P} & \textsc{S-P} & \textsc{DPrivLib} & \textsc{BinAgg} & \textsc{AdaSSP} & \textsc{Chen} & \textsc{Naive-L} & \textsc{Naive-G} & \textsc{Ada-P} & \textsc{S-P} & \textsc{DPrivLib} & \textsc{BinAgg} & \textsc{AdaSSP} & \textsc{Chen} & \textsc{Naive-L} & \textsc{Naive-G} \\
\midrule
1 & 220.420 & 220.420 & --- & 220.420 & 221.500 & 220.430 & 222.449 & \textbf{220.416} & \underline{220.419} & \underline{220.420} & --- & \underline{220.420} & \underline{220.420} & \underline{220.420} & 220.572 & \textbf{220.415} & \textbf{220.415} \\
2 & 216.290 & \textbf{216.290} & --- & \underline{216.300} & 219.200 & 216.360 & 219.328 & 216.303 & 216.312 & \textbf{216.290} & --- & \textbf{216.290} & 216.300 & \textbf{216.290} & 216.419 & \underline{216.293} & \underline{216.293} \\
3 & 209.110 & \textbf{209.110} & --- & \underline{209.130} & 213.160 & 209.260 & 213.030 & 209.142 & 209.155 & \textbf{209.110} & --- & \textbf{209.110} & 209.120 & \textbf{209.110} & 209.226 & \underline{209.111} & 209.112 \\
4 & 198.620 & \textbf{198.620} & --- & \underline{198.670} & 202.310 & 198.990 & 204.982 & 198.740 & 198.731 & \textbf{198.620} & --- & \textbf{198.620} & 198.640 & 198.630 & 198.841 & \underline{198.622} & 198.623 \\
5 & 187.710 & \textbf{187.720} & --- & \underline{187.850} & 192.490 & 188.380 & 197.884 & 187.978 & 187.904 & \textbf{187.710} & --- & \textbf{187.710} & 187.720 & 187.720 & 188.032 & 187.714 & \underline{187.712} \\
6 & 187.500 & \textbf{187.540} & \underline{187.560} & 187.880 & 194.210 & 188.660 & 201.615 & 188.213 & 187.906 & \textbf{187.500} & \textbf{187.500} & \textbf{187.500} & 187.530 & 187.510 & 188.022 & \underline{187.502} & \textbf{187.500} \\
8 & 182.753 & 183.575 & \textbf{183.151} & 185.110 & \underline{183.417} & 186.293 & 205.964 & 185.886 & 183.821 & \underline{182.762} & \textbf{182.757} & 182.775 & 182.81 & 182.798 & 183.967 & 182.786 & 182.767 \\
12 & 171.580 & --- & \textbf{176.790} & $2.67{\times}10^{6}$ & 184.100 & 183.360 & 212.651 & 213.923 & \underline{178.182} & --- & \textbf{171.630} & 171.810 & 172.200 & 171.970 & 175.395 & 171.946 & \underline{171.670} \\
15 & 161.074 & --- & \textbf{174.501} & $2.86{\times}10^{7}$ & 200.561 & 180.987 & 215.814 & 253.239 & \underline{177.916} & --- & \textbf{161.199} & 161.933 & 161.55 & 162.155 & 167.763 & 162.250 & \underline{161.279} \\
20 & 149.679 & --- & 212.164 & $1.12{\times}10^{8}$ & \underline{200.012} & \textbf{185.488} & 232.220 & 313.309 & 204.807 & --- & \textbf{150.040} & 154.355 & 150.60 & 153.125 & 160.898 & 154.512 & \underline{150.164} \\
24 & 141.180 & --- & 253.930 & $2.40{\times}10^{8}$ & \textbf{175.350} & \underline{190.090} & 248.854 & 349.747 & 213.555 & --- & \underline{142.080} & $1.55{\times}10^{5}$ & 143.15 & 147.950 & 156.920 & 163.900 & \textbf{141.951} \\
\bottomrule
\end{tabular*}
\end{table*}\endgroup

\begin{table*}[!t]
\color{black}
\captionsetup{font={color=black},labelfont={color=black}}
\centering
\caption{Correlated synthetic; $n=10{,}000p$}
\label{tab:correlated_combined}
\label{tab:synthetic_corr}
\label{tab:synthetic_corr_eps10}
\label{tab:corr_export1}
\label{tab:corr_export10}
\footnotesize
\setlength{\tabcolsep}{1.5pt}
\begin{tabular*}{\textwidth}{@{\extracolsep{\fill}}rr|rrrrrrrr|rrrrrrrr@{}}
\toprule
 & & \multicolumn{8}{c|}{$\varepsilon=1$} & \multicolumn{8}{c}{$\varepsilon=10$} \\
\cmidrule(lr){3-10}\cmidrule(lr){11-18}
$p$ & \textsc{OLS} & \textsc{Ada-P} & \textsc{S-P} & \textsc{DPrivLib} & \textsc{BinAgg} & \textsc{AdaSSP} & \textsc{Chen} & \textsc{Naive-L} & \textsc{Naive-G} & \textsc{Ada-P} & \textsc{S-P} & \textsc{DPrivLib} & \textsc{BinAgg} & \textsc{AdaSSP} & \textsc{Chen} & \textsc{Naive-L} & \textsc{Naive-G} \\
\midrule
4 & 0.0302 & \textbf{0.0303} & --- & \underline{0.0304} & 0.0328 & 0.0330 & 0.0413 & 0.0312 & 0.0309 & \textbf{0.0302} & --- & \textbf{0.0302} & \underline{0.0303} & \underline{0.0303} & 0.0310 & \underline{0.0303} & \underline{0.0303} \\
5 & 0.0299 & \textbf{0.0300} & --- & \underline{0.0303} & 0.0323 & 0.0318 & 0.0383 & 0.0322 & 0.0312 & \textbf{0.0299} & --- & \textbf{0.0299} & 0.0301 & \underline{0.0300} & 0.0305 & \textbf{0.0299} & \textbf{0.0299} \\
6 & 0.0300 & \textbf{0.0301} & \underline{0.0305} & 0.4455 & 0.0329 & 0.0322 & 0.0378 & 0.0329 & 0.0318 & \textbf{0.0300} & \textbf{0.0300} & \textbf{0.0300} & \underline{0.0301} & \textbf{0.0300} & 0.0304 & \textbf{0.0300} & \textbf{0.0300} \\
8 & 0.0303 & \textbf{0.0308} & \underline{0.0313} & 110.0000 & 0.0322 & 0.0314 & 0.0352 & 0.0338 & 0.0327 & \textbf{0.0303} & \textbf{0.0303} & \textbf{0.0303} & 0.0305 & \underline{0.0304} & 0.0306 & 0.0305 & \textbf{0.0303} \\
12 & 0.0303 & --- & \textbf{0.0314} & 320.0000 & 0.0377 & \underline{0.0315} & 0.0339 & 0.0361 & 0.0323 & --- & \textbf{0.0304} & 27.1000 & \textbf{0.0304} & \textbf{0.0304} & \underline{0.0305} & 0.0306 & \textbf{0.0304} \\
15 & 0.0300 & --- & \textbf{0.0310} & 950.0000 & 0.0446 & \underline{0.0316} & 0.0342 & 0.0387 & 0.0324 & --- & \textbf{0.0301} & 12.0000 & 0.0306 & \underline{0.0302} & 0.0303 & 0.0305 & \underline{0.0302} \\
20 & 0.0301 & --- & \underline{0.0319} & 2200.0000 & 0.0445 & \textbf{0.0308} & 0.0325 & 0.0418 & 0.0328 & --- & \textbf{0.0301} & 2.5000 & 0.0313 & \underline{0.0302} & \underline{0.0302} & 0.0306 & \underline{0.0302} \\
24 & 0.0300 & --- & 0.0332 & 5900.0000 & 0.0479 & \textbf{0.0312} & \underline{0.0328} & 0.0440 & 0.0330 & --- & \textbf{0.0302} & 39.7000 & 0.0327 & \textbf{0.0302} & \underline{0.0303} & 0.0313 & \underline{0.0303} \\
\bottomrule
\end{tabular*}
\end{table*}
\begin{table*}[!t]
\color{black}
\captionsetup{font={color=black},labelfont={color=black}}
\centering
\caption{i.i.d. synthetic test MSE ($\times 10^{-3}$); $n=10{,}000p$, $\varepsilon=1$.}
\label{tab:synthetic_iid_highp}
\footnotesize
\setlength{\tabcolsep}{4pt}
\begin{tabular*}{\textwidth}{@{\extracolsep{\fill}}rr|rrrrrrrr@{}}
\toprule
$p$ & \textsc{OLS} & \textsc{Ada-P} & \textsc{S-P} & \textsc{DPrivLib} & \textsc{BinAgg} & \textsc{AdaSSP} & \textsc{Chen} & \textsc{Naive-L} & \textsc{Naive-G} \\
\midrule
6 & 0.901 & \textbf{0.905} & \underline{0.911} & 0.918 & 1.050 & 1.000 & 2.464 & 1.011 & 0.949 \\
8 & 0.902 & \underline{0.929} & \textbf{0.923} & 0.933 & 1.090 & 1.000 & 2.314 & 1.080 & 0.956 \\
12 & 0.901 & --- & \textbf{0.954} & 0.981 & 3.570 & 0.998 & 2.230 & 1.305 & \underline{0.974} \\
15 & 0.899 & --- & \textbf{0.947} & 1.030 & 6.080 & 1.010 & 2.165 & 1.573 & \underline{0.985} \\
20 & 0.900 & --- & \textbf{0.988} & 1.190 & 8.840 & 1.040 & 2.136 & 2.470 & \underline{1.012} \\
24 & 0.899 & --- & \underline{1.040} & 1.380 & 9.270 & 1.060 & 2.102 & 3.711 & \textbf{1.029} \\
\bottomrule
\end{tabular*}
\end{table*}
\begingroup\color{black}
\section{S-PROBES: A Polynomial-Size Extension}
\label{sec:steiner_tmlr_main}

For large $p$, PROBES can be extended using Steiner-$(2,k,p+1)$\footnote{Every $2$-element subset of a ($p+1$)-element set appears in exactly one $k$-element block of the Steiner system.}
systems, for $k\in\{3,4,5\}$, by replacing the exponential-size
tensor with $O(p^2)$ constant-size simplex blocks on $k$ variables.
We call this extension S-PROBES;\footnote{We use the notation S-PROBES-$k$ to refer to the application of PROBES to each $k$-element block of the Steiner system.} it applies when the corresponding
Steiner system exists.

\paragraph{A seven-variable example.}
For $p=6$, every pair in $\{x_1,\ldots,x_6,y\}$ is contained in
exactly one block of the Steiner triple system
\[
\begin{aligned}
\mathcal T=\{&\{x_1,x_2,x_3\},\{x_1,x_4,x_5\},\{x_1,x_6,y\},\\
            &\{x_2,x_4,x_6\},\{x_2,x_5,y\},\{x_3,x_4,y\},\\
            &\{x_3,x_5,x_6\}\}.
\end{aligned}
\]
There are $21$ cross-terms: the $15$ sums
$\sum_i x_{j,i}x_{\ell,i}$ for $1\leq j<\ell\leq6$ and the six
sums $\sum_i x_{j,i}y_i$. A direct joint Laplace release of this
$21$-coordinate cross-term vector has $\ell_1$-sensitivity $21$
under add/remove adjacency on $[0,1]^7$. Using the full cross-term
budget $\varepsilon_1$ gives per-cross-term noise variance
$2\times21^2/\varepsilon_1^2$; no budget is divided among the terms.
The Steiner construction instead splits $\varepsilon_1$ across
seven blocks and applies the simplex construction within each block.
For example, the block $\{x_1,x_2,x_3\}$ partitions unity through
\[
1=((1-x_1)+x_1)((1-x_2)+x_2)((1-x_3)+x_3).
\]
Expanding the right-hand side gives eight nonnegative cells.
Each cross-term is recovered from two noisy cells, resulting in variance 
$4\times(7^2/\varepsilon_1^2)$ due to splitting of $\varepsilon_1$ across seven Steiner blocks. Here $\varepsilon_1$ covers the cross-term workload; the full OLS
mechanism must also account for the squared moments.

\paragraph{The S-PROBES-3 algorithm.}
Algorithm~\ref{alg:sprobes_full} gives the full pipeline for $p=6$.
The remaining construction details, proofs, variance analysis,
worked example, and detailed algorithms for S-PROBES-$4$ and S-PROBES-$5$ are in
Appendix~\ref{sec:steiner_tmlr_appendix}.

\paragraph{S-PROBES configurations used.}
For the S-PROBES entries in Tables~\ref{tab:covertype_low} and
\ref{tab:correlated_combined}, the supplied benchmark implementation
uses S-PROBES-3 for $p=6,8,12$, S-PROBES-4 for $p=15$, and
S-PROBES-5 for $p=20,24$, at both $\varepsilon=1$ and $10$.\footnote{The corresponding Steiner systems are $S(2,3,7)$, $S(2,3,9)$,
$S(2,3,13)$, $S(2,4,16)$, $S(2,5,21)$, and $S(2,5,25)$,
respectively; no dummy features are added.}

\paragraph{Correlated synthetic data generation.}
We partition the features into two blocks, setting
$q=\lfloor p/2\rfloor$. For the correlated block, we draw
$Z\in\mathbb{R}^{n\times q}$ and $A\in\mathbb{R}^{q\times q}$
with i.i.d.\ $\mathcal{N}(0,1)$ entries and construct
$X_{\mathrm{corr}}=\mathrm{MinMax}_{\mathrm{col}}(ZA^\top)$,
where column-wise min-max normalisation maps each feature
to $[0,1]$. The shared latent factors introduce dependence
among these features. The remaining $p-q$ features are drawn
i.i.d.\ $\mathrm{Uniform}(0,1)$, independently of $Z$ and $A$.
Responses follow the same clipped linear model as the IID
benchmark:
$y_i=\mathrm{clip}(0.2+\frac{0.6}{p}\sum_{j=1}^{p}x_{ij}
+\xi_i,0,1)$, with independent
$\xi_i\sim\mathcal{N}(0,0.03^2)$.
We use $p\in\{4,5,6,8,12,15,20,24\}$,
$n=10{,}000p$, and $\varepsilon\in\{1,10\}$,
with an 80/20 training/test split.
Table~\ref{tab:correlated_combined} reports mean per-fit
test RMSE over 100 seeds.

\paragraph{IID synthetic comparison.}
The IID data generator and 80/20 split are described in
Section~\ref{ssec:synthetic}. Table~\ref{tab:synthetic_iid_highp} extends the iid benchmark to
$p\in\{6,8,12,15,20,24\}$ at $n=10{,}000p$, with the same split
and $\varepsilon=1$.

Ada-PROBES is omitted at $p\geq 12$ where the tensor size becomes prohibitively large. Chen, AdaSSP and BinAgg have
approximate-DP parameters; train-derived preprocessing is not included
in the privacy accounting.

\paragraph{Baseline mechanisms and evaluation setup.}
Naive-L and Naive-G denote single, full-budget joint releases
of the $K=(p+1)(p+4)/2$ unique OLS sufficient-statistic sums, including
the private count. Under add/remove adjacency and fixed public bounds
and preprocessing into $[0,1]^{p+1}$, Naive-L adds independent
$\Lap(K/\varepsilon)$ noise to each coordinate and provides pure
$\varepsilon$-DP; Naive-G adds one analytically calibrated Gaussian
vector using joint $\ell_2$-sensitivity $\sqrt K$ and provides
$(\varepsilon,\delta)$-DP, where $\delta=n^{-2}$. Both use a spectral shift and ridge for the
OLS solve; Appendix~\ref{sec:naive_baselines} and
Algorithm~\ref{alg:naive_baselines} give the full construction.
These joint baselines release the complete OLS workload once, without
the Steiner representation.
Naive-L and S-PROBES provide pure $\varepsilon$-DP, whereas Naive-G
provides $(\varepsilon,\delta)$-DP.

Tables~\ref{tab:covertype_low} and~\ref{tab:correlated_combined}
report mean per-fit test RMSE at $\varepsilon=1,10$, using
100{,}000 training and 25{,}000 test records for Covertype and
$n=10{,}000p$ with an $80/20$ split for correlated synthetic data. Ada-P and S-P denote Ada-PROBES and S-PROBES, respectively.

Table~\ref{tab:covertype_low} also shows that \textsc{Ada-PROBES} achieves the lowest test RMSE on Covertype (D6) for $2 \le p \le 6$, while Naive-L and Naive-G have slightly lower RMSE at $p=1$.

\paragraph{S-PROBES discussion.}
On Covertype at $\varepsilon=1$, S-PROBES does not perform best
at every dimension (Table~\ref{tab:covertype_low}). At $p=20$, its
test RMSE is $212.16$, compared with $185.49$ for AdaSSP. At $p=24$,
its test RMSE is $253.93$, compared with $175.35$ for BinAgg.
At $\varepsilon=10$, S-PROBES has the lowest or
tied-lowest displayed private RMSE on Covertype at every evaluated
dimension $p=6,8,12,15,20$ (Table~\ref{tab:covertype_low}). At $p=24$,
the joint Gaussian mechanism has the advantage: its RMSE is $141.951$
versus S-PROBES's $142.080$, which remains within $1\%$ of OLS
($141.180$). On correlated data, S-PROBES has the lowest or tied-lowest
displayed private RMSE at $\varepsilon=10$ wherever it is available,
including $p=24$ ($0.0302$ versus joint Gaussian's $0.0303$;
Table~\ref{tab:correlated_combined}).
At $n=10{,}000p$ and $\varepsilon=1$, the higher-dimensional iid comparison
(Table~\ref{tab:synthetic_iid_highp}) shows S-PROBES with the lowest
displayed private MSE at $p=8,12,15,20$; Ada-P leads at $p=6$ and
the joint Gaussian baseline narrowly leads at $p=24$ ($1.029$ versus
$1.040$, in units of $10^{-3}$).  These results show that its performance depends on both the feature
dimension and the privacy budget. Thus, S-PROBES remains competitive
beyond the dimensions evaluated for Ada-PROBES, without leading at every $p$.

\paragraph{Limitation of S-PROBES when $p \ge 60$.}
For $\eps=1$, size parameter $n=10{,}000p$, and
$\delta=n^{-2}$, the analytically calibrated joint Gaussian release has lower noise variance for every sufficient-statistic sum than
S-\textsc{PROBES}-$k$, $k\in\{3,4,5\}$, at valid design dimensions
$p\ge60$ for diagonal groups of size $d=k-1$.
Appendix~\ref{sec:naive_baselines} supports this comparison with a
worked out example at $p=60$.
This exposes a high-dimensional variance disadvantage of these
Laplace-based S-\textsc{PROBES} constructions when approximate
$(\eps,\delta)$-DP is acceptable instead of pure $\eps$-DP.

\endgroup

\section{Conclusion}
\label{sec:conclusion}

We introduced \methodname{} (\textbf{PR}ivate \textbf{O}rdinary 
least squares via \textbf{BE}rnstein \textbf{S}implex), a 
single-shot $\varepsilon$-DP mechanism for multivariate linear 
regression on bounded data.
Our contribution is to specify the full OLS moment workload,
the variance of the DP noise in each of the recovered statistics, and a
noise-aware regularized solve for multivariate regression.
The Bernstein simplex embedding maps each record to a unit-$\ell_1$ vector with
$3^p \times 2$ entries, giving global $\ell_1$-sensitivity exactly $1$
under add/remove adjacency, regardless of $p$.
A single Laplace  vector  release at the full, unsplit budget $\varepsilon$ 
then recovers \emph{every} sufficient statistic for OLS as an 
unbiased linear post-processing step. Direct Laplace and Gaussian mechanisms also
make single joint releases; the distinction is PROBES's
Bernstein-simplex representation and its joint sensitivity, rather
than avoiding a per-statistic budget allocation in those releases.
This design is fundamentally distinct from the Bernstein 
Mechanism~\citep{alda2017bernstein} and the Functional 
Mechanism~\citep{zhang2012functional}: it targets sufficient-statistic 
privatisation rather than functional or loss perturbation.

  Our empirical evaluation on six real-world datasets and one synthetic
benchmark demonstrates that \textsc{Ada-PROBES} achieves the lowest
test RMSE across all evaluated privacy budgets and feature dimensions
on datasets D1, D4, and D5. On Covertype (D6), it achieves the lowest test RMSE for $2 \le p \le 6$; Naive-L and Naive-G have slightly lower RMSE at
$p=1$. \textsc{Ada-PROBES} also achieves the lowest test error in the low-dimensional
settings ($p \le 4$)
on datasets D2 and D3. However, exceptions arise on very small samples ($n = 10^3$ in D2) for $p \ge 5$ and $\varepsilon \le 1$, where \textsc{AdaSSP}
dominates over Ada-PROBES and achieves  lower error. Experimental results also indicate the failure of \textsc{DiffPrivLib} at low $\varepsilon$ (high privacy) regimes with $p\ge 5$, and
reinforce the importance of formal verification in privacy-critical
statistical software.

We also introduced S-PROBES, a polynomial-size
extension that replaces the exponential-size tensor with $O(p^2)$
constant-size simplex blocks using Steiner systems, where the
corresponding designs exist and we also evaluate it on one real world dataset (D6) and two synthetic datasets. S-PROBES retains pure $\varepsilon$-DP
and is evaluated for different feature dimensions up to $p=24$. At $\varepsilon=10$, S-PROBES has the lowest  RMSE on D6 at the evaluated dimensions from $p=6$ through $p=20$, while Naive-G achieves the lowest error for $p=24$. At $\varepsilon=1$ on D6, AdaSSP has the lowest RMSE for $p=20$ and BinAgg has lowest RMSE for $p=24$. For the synthetic dataset with correlated features, S-PROBES has the lowest RMSE on the evaluated dimensions up to $p=24$ at $\varepsilon=10$, while AdaSSP provides lowest RMSE for $p=20$ and $p=24$ at $\varepsilon=1$.

Several directions remain open.
A principled, privacy-compatible selection rule for the 
regularisation parameter $\lambda$ that accounts for the full 
correlated noise structure of the jointly released sufficient 
statistics (equation~\eqref{eq:opt_lambda}) would sharpen the performance.
Extending our proposed methods  to generalised linear models, quantile regression,
or ridge regression with unknown regularisation are natural next steps.

\begin{acks}
We thank the anonymous reviewers and the revision editor for their
detailed and constructive feedback throughout the review and revision
process, which substantially improved this paper.
This research received no specific grant from any funding agency in the public, commercial, or not-for-profit sectors.
\end{acks}

\begin{ai}
Generative AI-based tools were used during the preparation and
pre-submission review of the paper to assist with revising text,
improving clarity and flow, correcting grammatical errors and awkward
phrasing, and reviewing portions of the manuscript for consistency.
AI-based tools were also used to assist with code development,
debugging, and checking that aspects of the manuscript accurately
describe the experimental implementation.

AI-based tools were not used as part of the proposed research
methodology, to generate experimental data or results, or to generate
the figures or tables presented in the paper. The authors manually
reviewed and verified all AI-assisted output and are responsible for the
accuracy, originality, integrity, and content of the paper and its
associated research artifacts.
\end{ai}

\FloatBarrier

\FloatBarrier
\appendix
\renewcommand{\arraystretch}{0.82}

\section{Proofs for Section~\ref{subsec:bernstein}}
\label{proof:bernstein}

\subsection{Proof of Lemma~\ref{lem:bernstein}
(Bernstein Monomial Recovery)}

We wish to show that for $0 \le m \le 2d$ and $x \in [0,1]$,
\begin{equation*}
  x^m
  \;=\;
  \sum_{k=m}^{2d}
  \frac{\binom{k}{m}}{\binom{2d}{m}}\,
  \binom{2d}{k} x^k (1-x)^{2d-k}.
\end{equation*}

\begin{proof}
Starting from the right-hand side and simplifying the combined
binomial coefficient:
\begin{align*}
  \sum_{k=m}^{2d}
  \frac{\binom{k}{m}}{\binom{2d}{m}}
  \binom{2d}{k} x^k (1-x)^{2d-k}
  &\\
  {}=\frac{1}{\binom{2d}{m}}
  \sum_{k=m}^{2d}
  \binom{k}{m}\binom{2d}{k}
  \, x^k (1-x)^{2d-k}. &
\end{align*}
Apply the identity:
\begin{equation}
  \binom{k}{m}\binom{2d}{k}
  \;=\;
  \binom{2d}{m}\binom{2d-m}{k-m},
  \label{eq:subset_identity}
\end{equation}
which follows by expanding both sides as
$\frac{(2d)!}{m!\,(k-m)!\,(2d-k)!}$:
\begin{align*}
  &=
  \frac{\binom{2d}{m}}{\binom{2d}{m}}
  \sum_{k=m}^{2d}
  \binom{2d-m}{k-m}
  x^k (1-x)^{2d-k}\\
  &=
  \sum_{j=0}^{2d-m}
  \binom{2d-m}{j}
  x^{j+m} (1-x)^{2d-m-j}
  \qquad (j := k-m)\\
  &=
  x^m
  \sum_{j=0}^{2d-m}
  \binom{2d-m}{j}
  x^{j} (1-x)^{(2d-m)-j}\\
  &=
  x^m \cdot \bigl(x + (1-x)\bigr)^{2d-m}
  \quad\text{(binomial theorem)}\\
  &= x^m,
\end{align*}
which completes the proof.
\end{proof}

\section{The \textsc{Probes} Mechanism}
\label{sec:mechanism}

\subsection{Simple Linear Regression ($p=1$, $d=1$): Full Detail}
\label{ssec:case_a}

\subsubsection{Setup and Cells}
With $d=1$: $p_0(x)=(1-x)^2$, $p_1(x)=2x(1-x)$, $p_2(x)=x^2$.
Tensoring with $c_y(y)=y$ and $c_{\yb}(y)=1-y$ gives $3\times 2=6$ cells
(Table~\ref{tab:cells_a}), each non-negative and summing to 1 per record.

\begingroup\color{black}
\begin{table}[t]
\centering
\caption{All 6 cells of the Bernstein simplex transformation for $p=1$, $d=1$.}
\label{tab:cells_a}
\begin{tabular}{c l l}
\toprule
Cell & Per-record value & Dataset sum\\
\midrule
$\Bcell{2}{y}$   & $x_i^2\,y_i$             & $\sum_i x_i^2 y_i$\\
$\Bcell{2}{\yb}$ & $x_i^2\,(1-y_i)$         & $\sum_i x_i^2(1-y_i)$\\
$\Bcell{1}{y}$   & $2x_i(1-x_i)\,y_i$       & $\sum_i 2x_i(1-x_i)y_i$\\
$\Bcell{1}{\yb}$ & $2x_i(1-x_i)\,(1-y_i)$   & $\sum_i 2x_i(1-x_i)(1-y_i)$\\
$\Bcell{0}{y}$   & $(1-x_i)^2\,y_i$         & $\sum_i(1-x_i)^2 y_i$\\
$\Bcell{0}{\yb}$ & $(1-x_i)^2\,(1-y_i)$     & $\sum_i(1-x_i)^2(1-y_i)$\\
\bottomrule
\end{tabular}
\end{table}\endgroup

\subsubsection{Privacy}
$\varepsilon$-DP follows directly from the unit-norm
embedding (equation~\eqref{eq:unit_norm}) and the Laplace mechanism
(Proposition~\ref{prop:dp_facts}(i)), as proved in
\S\ref{sec:PROBES}; we do not re-derive it here.

\subsubsection{Inversion Weights and Recovery}
The recovery weights for $d=1$ are:
\begin{center}
\begin{tabular}{c|ccc}
\toprule
$m$ & $w_{m,0}$ & $w_{m,1}$ & $w_{m,2}$\\
\midrule
$0$ & $1$ & $1$ & $1$\\
$1$ & $0$ & $\tfrac{1}{2}$ & $1$\\
$2$ & $0$ & $0$ & $1$\\
\bottomrule
\end{tabular}
\end{center}

\begin{proposition}[Unbiased Estimators, $p=1$, $d=1$]
\begin{align}
  \hat{n} &= \tBcell{2}{y}+\tBcell{2}{\yb}+\tBcell{1}{y}+\tBcell{1}{\yb}+\tBcell{0}{y}+\tBcell{0}{\yb},\\
  \hat{S}_{x^2} &= \tBcell{2}{y}+\tBcell{2}{\yb},\\
  \hat{S}_{xy} &= \tBcell{2}{y}+\tfrac{1}{2}\tBcell{1}{y},\\
  \hat{S}_{x} &= \tBcell{2}{y}+\tBcell{2}{\yb}+\tfrac{1}{2}\tBcell{1}{y}+\tfrac{1}{2}\tBcell{1}{\yb},\\
  \hat{S}_{y} &= \tBcell{2}{y}+\tBcell{1}{y}+\tBcell{0}{y}.
\end{align}
Each is unbiased since $\E[Z_{k,r}]=0$.
\end{proposition}

\subsubsection{Structural Insight}
Pure $x$-moments (e.g.\ $\hat S_{x^2}$, $\hat S_x$) sum over \emph{both}
response components $r\in\{y,\yb\}$—the $y/(1-y)$ factors cancel via
$c_y+c_{\yb}=1$. Cross $xy$-moments (e.g.\ $\hat S_{xy}$) sum only
over $r=y$ cells, since $c_y(y)=y$ carries the response weight directly.
Consequently, cross-statistics use only the $r=y$ cells, whereas pure-$x$ statistics use both response components. At order $2$, this reduces the number of independent noise terms contributing to the cross-statistic and produces the variance factors reported in Table~\ref{tab:variances_a}.
Here $c_{kr}$ denotes the recovery weight on Bernstein cell $(k,r)$,
with coefficients ordered as
$(c_{2y},c_{2\yb},c_{1y},c_{1\yb},c_{0y},c_{0\yb})$.

\begingroup\color{black}
\begin{table}[t]
\centering
\caption{Estimation-error variances for $p=1$, $d=1$. $\sum c^2$ is the sum of squared
coefficients applied to independent noise terms of variance $2/\eps^2$.
\textbf{Key pattern:} $\Var(\hat S_{x^m y} - S_{x^m y}) = \tfrac{1}{2}\Var(\hat S_{x^m} - S_{x^m})$
because cross-statistics use only $r=y$ cells.}
\label{tab:variances_a}
\setlength{\tabcolsep}{3pt}
\renewcommand{\arraystretch}{1.15}
\begin{tabular*}{\linewidth}{@{\extracolsep{\fill}} l c c c @{}}
\toprule
Statistic &
  \makecell[c]{Recovery coeffs.$(c_{kr})$} &
  $\sum c^2$ &
  \makecell[c]{$\mathrm{Var}(\hat{\cdot}-\cdot)$$\times\,\eps^{-2}$} \\
\midrule
$\hat{n}$       & $(1,1,1,1,1,1)$                       & $6$             & $12$\\
$\hat{S}_{x^2}$ & $(1,1,0,0,0,0)$                       & $2$             & $4$\\
$\hat{S}_{xy}$  & $(1,0,\tfrac{1}{2},0,0,0)$             & $\tfrac{5}{4}$  & $\tfrac{5}{2}$\\
$\hat{S}_x$     & $(1,1,\tfrac{1}{2},\tfrac{1}{2},0,0)$   & $\tfrac{5}{2}$  & $5$\\
$\hat{S}_y$     & $(1,0,1,0,1,0)$                       & $3$             & $6$\\
\bottomrule
\end{tabular*}
\end{table}\endgroup

\begingroup\color{black}
\begin{algorithm}[t]
\caption{\methodname{}: Simple Linear Regression ($p=1$, $d=1$)}
\label{alg:dp_bsm_a}
\SetKwInOut{Input}{Input}
\SetKwInOut{Privacy}{Privacy}
\SetKwInOut{Output}{Output}
\Input{$D=\{(x_i,y_i)\}_{i=1}^n\subset[0,1]^2$}
\Privacy{$\eps>0$ \emph{(single, unsplit budget)}}
\Output{Private OLS coefficients $(\hat\alpha,\hat\beta)$}
\BlankLine
Initialise $\Bcell{k}{r}\gets 0$ for all $(k,r)$\;
\For{each $(x_i,y_i)\in D$}{
  $\Bcell{2}{y} \mathrel{+}= x_i^2\,y_i$;\quad
  $\Bcell{2}{\yb} \mathrel{+}= x_i^2\,(1-y_i)$\;
  $\Bcell{1}{y} \mathrel{+}= 2x_i(1-x_i)\,y_i$;\quad
  $\Bcell{1}{\yb} \mathrel{+}= 2x_i(1-x_i)\,(1-y_i)$\;
  $\Bcell{0}{y} \mathrel{+}= (1-x_i)^2\,y_i$;\quad
  $\Bcell{0}{\yb} \mathrel{+}= (1-x_i)^2\,(1-y_i)$\;
}
\For{each $(k,r)\in\{0,1,2\}\times\{y,\yb\}$}{
  $\tBcell{k}{r} \gets \Bcell{k}{r} + \Lap(0,\,1/\eps)$\;
}
$\hat{n} \gets \sum_{k,r}\tBcell{k}{r}$;\quad
\lIf{$\hat{n}\leq 0$}{\Return $(0,\;0.5)$}
$\hat{S}_{x^2} \gets \tBcell{2}{y}+\tBcell{2}{\yb}$;\quad
$\hat{S}_{xy} \gets \tBcell{2}{y}+\tfrac{1}{2}\tBcell{1}{y}$\;
$\hat{S}_{x} \gets \tBcell{2}{y}+\tBcell{2}{\yb}+\tfrac{1}{2}\tBcell{1}{y}+\tfrac{1}{2}\tBcell{1}{\yb}$\;
$\hat{S}_{y} \gets \tBcell{2}{y}+\tBcell{1}{y}+\tBcell{0}{y}$\;
$\Delta \gets \hat{S}_{x^2}\,\hat{n} - \hat{S}_x^2$;\quad
\lIf{$\Delta\leq 0$}{\Return $(0,\;0.5)$}
$\hat\alpha \gets (\hat{n}\,\hat{S}_{xy} - \hat{S}_x\,\hat{S}_y)/\Delta$;\quad
$\hat\beta \gets (\hat{S}_{x^2}\,\hat{S}_y - \hat{S}_x\,\hat{S}_{xy})/\Delta$\;
\Return $(\hat\alpha,\hat\beta)$\;
\end{algorithm}\endgroup

\subsection{General Framework: $p$ Bounded Features, Degree $d$}
\label{ssec:general_probes}

\begingroup\color{black}
\begin{algorithm}[t]
\caption{\methodname{}: General framework for $p$ features,
         degree $d \ge 1$ (pure $\varepsilon$-DP)}
\label{alg:dp_bsm_general}
\SetKwInOut{Input}{Input}
\SetKwInOut{Privacy}{Privacy}
\SetKwInOut{Output}{Output}
\Input{$D=\{(\mathbf{x}_{i},y_i)\}_{i=1}^n\subset[0,1]^{p+1}$; degree $d\geq 1$}
\Privacy{$\eps>0$}
\Output{Privatised sufficient statistics: $\hat{n}$,
        $\{\hat{S}_{\mathbf{m}}\}$, $\{\hat{S}_{\mathbf{m},y}\}$}
\BlankLine
$p_{k_s}(x) \gets \binom{2d}{k_s} x^{k_s}(1-x)^{2d-k_s}$ for $k_s \in \{0,\ldots,2d\}$\;
$c_y(y) \gets y$,\quad $c_{1-y}(y) \gets 1-y$\;
$\Bcell{\bk}{r} \gets 0$ for all $\bk \in \{0,\ldots,2d\}^p$, $r \in \{y, 1-y\}$\;
\For{each record $(\mathbf{x}_i,y_i) \in D$}{
  \For{each $\bk \in \{0,\ldots,2d\}^p$ and $r \in \{y, 1-y\}$}{
    $\Bcell{\bk}{r} \gets \Bcell{\bk}{r} + \Bigl(\prod_{s=1}^{p} p_{k_s}(x_{is})\Bigr)\cdot c_r(y_i)$\;
  }
}
\For{each $(\bk, r)$}{
  Draw $Z_{\bk,r} \overset{\mathrm{iid}}{\sim} \mathrm{Lap}(1/\eps)$\;
  $\tBcell{\bk}{r} \gets \Bcell{\bk}{r} + Z_{\bk,r}$\;
}
$\hat{n}\gets\displaystyle\sum_{\bk,r}\tBcell{\bk}{r}$\;
\For{each multi-index $\mathbf{m} \in \{0,\ldots,2d\}^p$}{
  $\hat{S}_{\mathbf{m}}\gets
    \displaystyle\sum_{\bk,\,r}
    \Bigl(\prod_{s=1}^{p} w_{m_s,k_s}\Bigr)\,\tBcell{\bk}{r}$\;
  $\hat{S}_{\mathbf{m},y}\gets
    \displaystyle\sum_{\bk}
    \Bigl(\prod_{s=1}^{p} w_{m_s,k_s}\Bigr)\,\tBcell{\bk}{y}$\;
}
Assemble $\hat{A}\in\mathbb{R}^{(p+1)\times(p+1)}$ and $\hat{\mathbf{b}}\in\mathbb{R}^{p+1}$ from $\hat{n}$, $\{\hat{S}_{\mathbf{m}}\}$, $\{\hat{S}_{\mathbf{m},y}\}$\;
$\rho \gets (2/\eps)\cdot(2d+1)^{(p-1)/2}$\;
$\lambda_{\min}^{-} \gets \min\bigl(\lambda_{\min}(\hat{A}),\,0\bigr)$\;
$\widetilde{\boldsymbol{\theta}} \gets (\hat{A}-\lambda_{\min}^{-}I+\rho I)^{-1}\hat{\mathbf{b}}$\;
\Return{$\widetilde{\boldsymbol{\theta}}$}\;
\end{algorithm}\endgroup

\begin{definition}[Tensor Bernstein Simplex Map]
For $d\geq 1$, $p\geq 1$, map each record $(x_{i1},\ldots,x_{ip},y_i)\in[0,1]^{p+1}$ to
the $N=(2d+1)^p\times 2$ dimensional vector
\[
\begin{aligned}
  \mathbf{u}_i((\bk,r))
  &= \Bigl(\prod_{s=1}^p p_{k_s}(x_{is})\Bigr)c_r(y_i),\\
  &\qquad \bk\in\{0,\ldots,2d\}^p,\quad r\in\{y,\yb\}.
\end{aligned}
\]
One verifies $\|\mathbf{u}_i\|_1=1$ since $\sum_k p_k(x)=1$ and $c_y+c_{\yb}=1$.
\end{definition}
\begin{theorem}[General Estimators and Estimation-Error Variances]
\label{thm:general}
For multi-index $\bm=(m_1,\ldots,m_p)$ with $m_s\leq 2d$:
\begin{align}
  \hat{S}_{\mathbf{x}^\mathbf{m}} &\coloneqq
    \sum_{\bk,r}\Bigl(\prod_s w_{m_s,k_s}\Bigr)\tBcell{\bk}{r},\\
  \hat{S}_{\mathbf{x}^{\mathbf{m}'}y} &\coloneqq
    \sum_{\bk}\Bigl(\prod_s w_{m'_s,k_s}\Bigr)\tBcell{\bk}{y}.
\end{align}
Both are unbiased, with estimation-error variances:
\begin{align}
  \Var\bigl(\hat{S}_{\mathbf{x}^\mathbf{m}} - S_{\mathbf{x}^\mathbf{m}}\bigr) &= \frac{4}{\eps^2}\prod_{s=1}^p\sigma_{m_s,d}^2,\\
  \Var\bigl(\hat{S}_{\mathbf{x}^{\mathbf{m}'}y} - S_{\mathbf{x}^{\mathbf{m}'}y}\bigr) &= \frac{2}{\eps^2}\prod_{s=1}^p\sigma_{m'_s,d}^2,\\
  \Var(\hat{n} - n) &= \frac{4(2d+1)^p}{\eps^2}.
\end{align}
\end{theorem}

The Bernstein construction extends immediately to degree-$d$ polynomial
regression by using the order-$2d$ basis, yielding a simplex of
$(2d+1)^p\times2$ cells. The per-record $\ell_1$-sensitivity remains $1$,
so a single Laplace release at scale $1/\eps$ gives $\eps$-DP.

\noindent Variance expressions for general $d$ follow
the same pattern as Theorem~\ref{thm:general} with $\sigma_{m,1}^2$
replaced by $\sigma_{m,d}^2$; see Definition~\ref{def:variance_factor}.

\subsection{Regularisation}
\label{sec:regularisation}

At high privacy ($\eps\leq0.5$) or large $p$, the noisy Gram matrix
$\hat A$ recovered via the PROBES estimators may fail to be positive
definite. We apply a   regulariser: let $\lambda:=\min(\lambda_{\min}(\hat A),0)$
and $\rho:=(2/\eps)\cdot 3^{(p-1)/2}$. The private OLS estimator is
\begin{equation}
\widetilde{\boldsymbol\theta}
:= (\hat A - \lambda I + \rho I)^{-1}\hat{\mathbf{b}}.
\end{equation}
which remains $\varepsilon$-DP by post-processing immunity (Proposition~\ref{prop:dp_facts}(ii)).

\begingroup\color{black}
\section{S-PROBES: Supporting Construction and Analysis}
\label{sec:steiner_tmlr_appendix}
\subsection{Steiner-Block Framework}
\label{sec:steiner_framework}

\begin{definition}[Steiner System $S(2,k,v)$]
\label{def:steiner}
A \emph{Steiner system} $S(2,k,v)$ is a collection of $k$-element
subsets (\emph{blocks}) of a $v$-element ground set such that every
$2$-element subset of the ground set appears in \emph{exactly one} block.
\end{definition}

\paragraph{Existence of Steiner systems.}
The necessary and sufficient conditions for the existence of $S(2,k,v)$
are classical combinatorial results.
For $k=3$: $v \equiv 1$ or $3 \pmod{6}$~\citep{kirkman1847}.
For $k=4$ and $k=5$: the necessary conditions
$v \equiv 1$ or $4 \pmod{12}$ and $v \equiv 1$ or $5 \pmod{20}$,
respectively, are also \emph{sufficient}, as proved by \citet{hanani1961}.
In our context $v = p+1$, so the conditions on $p$ are:
$S(2,3,p{+}1)$: $p \equiv 0$ or $2 \pmod{6}$;
$S(2,4,p{+}1)$: $p \equiv 0$ or $3 \pmod{12}$;
$S(2,5,p{+}1)$: $p \equiv 0$ or $4 \pmod{20}$.

\noindent
We apply $S(2,k,p+1)$ to the $p+1$ variables
$\{x_1,\ldots,x_p,y\}$, so every pair $(x_j,x_\ell)$ and $(x_j,y)$
is covered by exactly one block.
This matches the structure of the Gram matrix: each entry
$S_{x_j x_\ell}$ or $S_{x_j y}$ is privatised by exactly one block
with no redundancy.

\paragraph{Three block sizes.}
Table~\ref{tab:sprobes_variants} summarises the three S-\textsc{PROBES}-$k$
variants, corresponding to $k \in \{3,4,5\}$.
Larger $k$ yields fewer blocks and hence smaller total sensitivity $M$,
at the cost of a larger per-block simplex dimension ($2^k$ cells per block).
The \emph{selection guideline} is: use S-\textsc{PROBES}-3 for
$6 \le p \le 11$, S-\textsc{PROBES}-4 for $12 \le p \le 19$, and
S-\textsc{PROBES}-5 for $p \ge 20$.
If $p$ does not satisfy the existence condition for the chosen $k$,
pad with zero-valued dummy features to the nearest valid $p$ and discard
the dummy coefficients after recovery.

\begingroup\color{black}
\begin{table}[t]
\color{black}
\captionsetup{font={color=black},labelfont={color=black}}

\centering
\caption{S-\textsc{PROBES}-$k$ variant comparison.
$M$ is the total block count and equals the $\ell_1$-sensitivity.
The counts assume an existing Steiner design and a partition of the
$p$ features into groups of size $k-1$, so $(k-1)\mid p$;
the listed congruences imply this divisibility for $k=3,4,5$.
Existence conditions follow from \citet{kirkman1847} ($k{=}3$) and
\citet{hanani1961} ($k{=}4,5$).}
\label{tab:sprobes_variants}
\setlength{\tabcolsep}{3pt}
\begin{tabular}{@{}c l c c c@{}}
\toprule
$k$ & System & \makecell[c]{$p$\\condition}
    & \makecell[c]{$B$\\blocks} & \makecell[c]{$M$\\total} \\
\midrule
$3$ & $S(2,3,p{+}1)$ & $p \equiv 0,2 \!\pmod{6}$
    & $\frac{p(p+1)}{6}$ & $\frac{p(p+4)}{6}$ \\[4pt]
$4$ & $S(2,4,p{+}1)$ & $p \equiv 0,3 \!\pmod{12}$
    & $\frac{p(p+1)}{12}$ & $\frac{p(p+5)}{12}$ \\[4pt]
$5$ & $S(2,5,p{+}1)$ & $p \equiv 0,4 \!\pmod{20}$
    & $\frac{p(p+1)}{20}$ & $\frac{p(p+6)}{20}$ \\
\bottomrule
\end{tabular}
\end{table}\endgroup

\subsection{Block Embedding and Unit Sensitivity}
\label{sec:block_embedding}

\paragraph{Steiner blocks.}
For any $k$, each Steiner block 
$\{z_1,\ldots,z_k\} \subseteq
\{x_1,\ldots,x_p,y\}$ is mapped to a $2^k$-dimensional binary
product-simplex vector via the tensor expansion
\[
  \bigl[(1-z_1)+z_1\bigr]\otimes\bigl[(1-z_2)+z_2\bigr]
  \otimes\cdots\otimes\bigl[(1-z_k)+z_k\bigr],
\]
giving all $2^k$ products
$\prod_{i=1}^{k} z_i^{u_i}(1-z_i)^{1-u_i}$,
$(u_1,\ldots,u_k)\in\{0,1\}^k$,
whose $\ell_1$-norm equals exactly $1$.

\paragraph{Diagonal blocks.}
Squared moments $S_{x_j^2}$ are not covered by the Steiner blocks
(which only touch each variable's \emph{linear} contribution in the
product).
Groups of $k-1$ features are therefore mapped to a $2^{k-1}$-dimensional
simplex using $x_j^2$ instead of $x_j$, again with $\ell_1$-norm $1$.
This construction assumes $(k-1)\mid p$ and a disjoint partition of all
$p$ features into $D=p/(k-1)$ diagonal blocks. Together with an existing
$S(2,k,p+1)$ design, this gives
\[
 B=\frac{p(p+1)}{k(k-1)},\qquad
 M=B+D=\frac{p(p+k+1)}{k(k-1)}.
\]
If divisibility fails, these counts do not apply unchanged: use a smaller
final diagonal block or pad with zero-valued dummy features to a dimension
satisfying both design existence and divisibility, and recalibrate using
the actual total block count.

\paragraph{Total sensitivity.}
Adding or removing any single record changes each block's aggregate
by at most $1$ in $\ell_1$-norm.
Because there are $M$ blocks in total, the global $\ell_1$-sensitivity
of the joint release is exactly $M$.

\subsection{Privacy Guarantee}
\label{sec:sprobes_privacy}

\begin{theorem}[$\eps$-DP of S-\textsc{PROBES}]
\label{thm:sprobes_dp}
The S-\textsc{PROBES} mechanism described above satisfies
$\eps$-differential privacy under add/remove adjacency, assuming the
Steiner design and the diagonal partition described above exist.
\end{theorem}

\begin{proof}
Adding or removing record $i$ changes:
\begin{itemize}[nosep]
  \item Each Steiner block $B_\ell$: the $2^k$-cell vector
        $\phi^{(B_\ell)}_i$ satisfies $\|\phi^{(B_\ell)}_i\|_1=1$.
  \item Each diagonal block $D_m$: the $2^{k-1}$-cell vector
        $\psi^{(D_m)}_i$ satisfies $\|\psi^{(D_m)}_i\|_1=1$.
\end{itemize}
Total $\ell_1$-sensitivity is therefore
\[
 M=B+D=\frac{p(p+k+1)}{k(k-1)},
 \qquad D=\frac{p}{k-1},
\]
under $(k-1)\mid p$. For $k=3$, this reduces to
$M=B+p/2=p(p+4)/6$.
The Laplace mechanism at scale $M/\eps$ gives $\eps$-DP
(Proposition~\ref{prop:dp_facts}(i)).
All downstream steps like linear combinations, averaging, matrix
assembly, and the regularised solve are deterministic
post-processing and preserve $\eps$-DP by
Proposition~\ref{prop:dp_facts}(ii).
\end{proof}

\subsection{Variance Analysis}
\label{sec:sprobes_variance}

Each Laplace draw satisfies $\Var(Z) = 2M^2/\eps^2$.
Table~\ref{tab:probes_vs_sprobes_p6} gives per-statistic estimation-error variances
for $p=6$ ($M=10$, $r=3$ Steiner blocks per variable).

\begingroup\color{black}
\begin{table*}[t]
\color{black}
\captionsetup{font={color=black},labelfont={color=black}}

\centering
\caption{\textsc{PROBES} ($N=3^6{\times}2=1458$ cells, scale $1/\eps$)
vs.\ S-\textsc{PROBES}-3 ($M=10$ blocks, scale $10/\eps$) at $p=6$.
All estimation-error variances $\mathrm{Var}(\hat{\cdot}-\cdot)$, $\times\,\eps^2$.
Cross-products and squared moments share the same S-\textsc{PROBES}
estimation-error variance $4M^2$; linear moments and $\hat n$ benefit
from $r$-fold averaging ($r=3$).}
\label{tab:probes_vs_sprobes_p6}
\setlength{\tabcolsep}{4pt}
\begin{tabular}{lccccc}
\toprule
Statistic
  & \makecell[c]{Noisy cells\\used (S-P)}
  & \makecell[c]{$\mathrm{Var}_{\text{S-P}}$\\formula}
  & $\mathrm{Var}_\text{P}(\hat{\cdot}-\cdot)\cdot\eps^2$
  & $\mathrm{Var}_\text{S-P}(\hat{\cdot}-\cdot)\cdot\eps^2$
  & Ratio P/\,S-P \\
\midrule
$\hat S_{x_j x_\ell}$
  & 2 from 1 Steiner block & $4M^2$
  & $506.25$ & $400.00$ & $1.27\times$ \\
$\hat S_{x_j y}$
  & 2 from 1 Steiner block & $4M^2$
  & $607.50$  & $400.00$ & $1.52\times$ \\
$\hat S_{x_j^2}$
  & 2 from 1 diagonal block & $4M^2$
  & $972.00$    & $400.00$ & $2.43\times$ \\
$\hat S_{x_j}$
  & $4{\times}3$ Steiner cells & $8M^2/3$
  & $1215.00$   & $\approx267.00$ & $4.56\times$ \\
$\hat S_{y}$
  & $4{\times}3$ Steiner cells & $8M^2/3$
  & $1458.00$   & $\approx267.00$ & $5.47\times$ \\
$\hat n$
  & $4{\times}3$ diagonal cells & $8M^2/3$
  & $2916.00$   & $\approx267.00$ & $10.94\times$ \\
\bottomrule
\end{tabular}
\end{table*}\endgroup

\paragraph{Asymptotic advantage.}
For the $k=3$ construction, $M = p(p+4)/6 = \Theta(p^2)$, so S-\textsc{PROBES} estimation-error variances grow as
$\Theta(p^4)$, while \textsc{PROBES} estimation-error variances grow as $\Theta(3^p)$:
\[
  \frac{\Var_P}{\Var_{SP}}
  = \Theta\!\left(\frac{3^p}{p^4}\right)
  \;\xrightarrow{p\to\infty}\; \infty.
\]
S-\textsc{PROBES} gains an exponentially growing advantage as $p$
increases, making it a tractable \textsc{PROBES} variant
for $p \ge 12$.

\subsection{S-\textsc{Probes}-3: Case $p=6$ (Fano Plane)}
\label{sec:p6}

The Fano plane is the unique Steiner triple system $S(2,3,7)$:
it covers all $\binom{7}{2}=21$ pairs of 7 points in $7$ triples.

\subsubsection{Block Structure}

\begin{definition}[Fano Blocks for $p=6$]
Fix the mapping $x_j\mapsto j$ ($j=1,\ldots,6$), $y\mapsto7$.
The 7 Steiner blocks are:
\begin{center}
\begin{tabular}{cl}
\toprule
Block & Variables\\
\midrule
$B_1=\{1,2,4\}$ & $\{x_1,x_2,x_4\}$\\
$B_2=\{2,3,5\}$ & $\{x_2,x_3,x_5\}$\\
$B_3=\{3,4,6\}$ & $\{x_3,x_4,x_6\}$\\
$B_4=\{4,5,7\}$ & $\{x_4,x_5,y\}$\\
$B_5=\{1,5,6\}$ & $\{x_1,x_5,x_6\}$\\
$B_6=\{2,6,7\}$ & $\{x_2,x_6,y\}$\\
$B_7=\{1,3,7\}$ & $\{x_1,x_3,y\}$\\
\bottomrule
\end{tabular}
\end{center}
Each variable appears in exactly $r=p/2=3$ blocks. Every pair appears in exactly one block.
\end{definition}

\begin{definition}[Diagonal Blocks for $p=6$]
The 3 diagonal blocks pair the 6 bounded features:
$D_1=\{x_1,x_2\}$, $D_2=\{x_3,x_4\}$, $D_3=\{x_5,x_6\}$.
Each maps to a 4-dimensional simplex:
\[
  \psi_i^{(D_k)} = \begin{pmatrix}
    (1-x_{i,2k-1}^2)(1-x_{i,2k}^2)\\
    (1-x_{i,2k-1}^2)x_{i,2k}^2\\
    x_{i,2k-1}^2(1-x_{i,2k}^2)\\
    x_{i,2k-1}^2 x_{i,2k}^2
  \end{pmatrix},
  \quad \|\psi_i^{(D_k)}\|_1=1.
\]
\end{definition}

Total block count: $M=B+p/2=7+3=10$.

\subsection{S-\textsc{Probes}-4 and S-\textsc{Probes}-5}
\label{sec:sprobes45}

\begingroup\color{black}
\begin{algorithm}[t]
\color{black}
\SetAlCapFnt{\color{black}}
\SetAlCapNameFnt{\color{black}}
\SetAlFnt{\small}
\SetNlSty{red}{}{}

\caption{S-\textsc{Probes}-4 ($p\geq12$, $p\equiv0,3\pmod{12}$)}
\label{alg:sprobes4}
\SetKwInOut{Input}{Input}
\SetKwInOut{Output}{Output}
\Input{$p$ features; $\{(x_{i,1},\ldots,x_{i,p},y_i)\}_{i=1}^n$; $\eps>0$}
\Output{Privatised block sums for recovery of all sufficient statistics}
\BlankLine
$B\gets p(p+1)/12$;\quad $M\gets p(p+5)/12$\;
Construct $S(2,4,p+1)$: blocks $\{B_1,\ldots,B_B\}$, each a 4-element
  subset of $\{x_1,\ldots,x_p,y\}$ covering every pair exactly once\;
Partition features into $p/3$ diagonal groups $\{D_1,\ldots,D_{p/3}\}$
  of size $3$\;
\BlankLine
\For{each Steiner block $B_\ell=\{a,b,c,d\}$}{
  $\begin{aligned}
    \Phi^{(B_\ell)}_{uvwt}\gets\sum_i &a_i^u(1{-}a_i)^{1-u}\,b_i^v(1{-}b_i)^{1-v}\\
      &{}\cdot c_i^w(1{-}c_i)^{1-w}\,d_i^t(1{-}d_i)^{1-t}
  \end{aligned}$
  \quad $\forall\,(u,v,w,t)\in\{0,1\}^4$\;
  Release $\tilde\Phi^{(B_\ell)}_{uvwt}\gets\Phi^{(B_\ell)}_{uvwt}+Z,\;
    Z\overset{\mathrm{iid}}{\sim}\Lap(M/\eps)$\;
}
\For{each diagonal group $D_k=\{x_{3k-2},x_{3k-1},x_{3k}\}$}{
  $\begin{aligned}
    \Phi^{(D_k)}_{uvw}\gets\sum_i &(x_{i,3k-2}^2)^u(1{-}x_{i,3k-2}^2)^{1-u}\\
      &{}\cdot (x_{i,3k-1}^2)^v(1{-}x_{i,3k-1}^2)^{1-v}\\
      &{}\cdot (x_{i,3k}^2)^w(1{-}x_{i,3k}^2)^{1-w}
  \end{aligned}$
  \quad $\forall\,(u,v,w)\in\{0,1\}^3$\;
  Release $\tilde\Phi^{(D_k)}_{uvw}\gets\Phi^{(D_k)}_{uvw}+Z,\;
    Z\overset{\mathrm{iid}}{\sim}\Lap(M/\eps)$\;
}
\Return{$\{\tilde\Phi^{(B_\ell)}\}_{\ell=1}^{B}$, $\{\tilde\Phi^{(D_k)}\}_{k=1}^{p/3}$}\;
\end{algorithm}\endgroup

\paragraph{Recovery and parameters (S-\textsc{PROBES}-4).} Algorithm~\ref{alg:sprobes4} details differentially private release of Steiner blocks and diagonal blocks in S-PROBES-$4$. Recovery of $\hat S_{x_jx_\ell}$, $\hat S_{x_jy}$, $\hat S_{x_j^2}$,
$\hat S_{x_j}$, $\hat S_y$, and $\hat n$ from the released blocks
follows the recovery pattern illustrated in Algorithm~\ref{alg:sprobes_full}
(marginalising over the indicator coordinates corresponding to each
variable, and averaging over the $r = p/3$ Steiner blocks in which
each feature appears), and is omitted here for brevity.
The assumed congruence $p\equiv0,3\pmod{12}$ ensures $3\mid p$.
There are $B = p(p+1)/12$ Steiner blocks and $D=p/3$
disjoint diagonal groups of size $3$, with each feature appearing in
$r = p/3$ Steiner blocks and total sensitivity $M = p(p+5)/12$.

\begingroup\color{black}
\begin{algorithm}[t]
\color{black}
\SetAlCapFnt{\color{black}}
\SetAlCapNameFnt{\color{black}}
\SetAlFnt{\small}
\SetNlSty{red}{}{}

\caption{S-\textsc{Probes}-5 ($p\geq20$, $p\equiv0,4\pmod{20}$)}
\label{alg:sprobes5}
\SetKwInOut{Input}{Input}
\SetKwInOut{Output}{Output}
\Input{$p$ features; $\{(x_{i,1},\ldots,x_{i,p},y_i)\}_{i=1}^n$; $\eps>0$}
\Output{Privatised block sums for recovery of all sufficient statistics}
\BlankLine
$B\gets p(p+1)/20$;\quad $M\gets p(p+6)/20$\;
Construct $S(2,5,p+1)$: blocks $\{B_1,\ldots,B_B\}$, each a 5-element
  subset of $\{x_1,\ldots,x_p,y\}$ covering every pair exactly once\;
Partition features into $p/4$ diagonal groups $\{D_1,\ldots,D_{p/4}\}$
  of size $4$\;
\BlankLine
\For{each Steiner block $B_\ell=\{a,b,c,d,e\}$}{
  $\Phi^{(B_\ell)}_{uvwtq}\gets\sum_i a_i^u(1{-}a_i)^{1-u}\cdots e_i^q(1{-}e_i)^{1-q}$
  \quad $\forall\,(u,v,w,t,q)\in\{0,1\}^5$\;
  Release $\tilde\Phi^{(B_\ell)}_{uvwtq}\gets\Phi^{(B_\ell)}_{uvwtq}+Z,\;
    Z\overset{\mathrm{iid}}{\sim}\Lap(M/\eps)$\;
}
\For{each diagonal group $D_k=\{x_{4k-3},\ldots,x_{4k}\}$}{
  $\begin{aligned}
    \Phi^{(D_k)}_{uvwt}\gets\sum_i &(x_{i,4k-3}^2)^u(1{-}x_{i,4k-3}^2)^{1-u}\\
      &{}\cdots (x_{i,4k}^2)^t(1{-}x_{i,4k}^2)^{1-t}
  \end{aligned}$
  \quad $\forall\,(u,v,w,t)\in\{0,1\}^4$\;
  Release $\tilde\Phi^{(D_k)}_{uvwt}\gets\Phi^{(D_k)}_{uvwt}+Z,\;
    Z\overset{\mathrm{iid}}{\sim}\Lap(M/\eps)$\;
}
\Return{$\{\tilde\Phi^{(B_\ell)}\}_{\ell=1}^{B}$, $\{\tilde\Phi^{(D_k)}\}_{k=1}^{p/4}$}\;
\end{algorithm}\endgroup

\paragraph{Recovery and parameters (S-\textsc{PROBES}-5).}
Algorithm~\ref{alg:sprobes5} details differentially private release of Steiner blocks and diagonal blocks in S-PROBES-$5$. Recovery follows the pattern illustrated in Algorithm~\ref{alg:sprobes_full},
marginalising over the relevant indicator coordinates and averaging
over the $r = p/4$ Steiner blocks containing each feature.
The assumed congruence $p\equiv0,4\pmod{20}$ ensures $4\mid p$.
There are $B = p(p+1)/20$ Steiner blocks and $D=p/4$
disjoint diagonal groups of size $4$, with total sensitivity $M = p(p+6)/20$.

\paragraph{Comparison across block sizes.}
Increasing the Steiner block size $k$ trades fewer total blocks for a
larger per-block simplex, reducing the overall sensitivity $M$.
At $p=12$, S-\textsc{PROBES}-4 gives
$M = \tfrac{12 \cdot 17}{12} = 17$, versus S-\textsc{PROBES}-3's
$M = \tfrac{12 \cdot 16}{6} = 32$ --- a $1.88\times$ reduction in
noise scale.
The advantage of larger $k$ grows with $p$: at $p=20$,
S-\textsc{PROBES}-5 gives $M = \tfrac{20 \cdot 26}{20} = 26$, versus
S-\textsc{PROBES}-3's $M = \tfrac{20 \cdot 24}{6} = 80$, a
$3.1\times$ reduction.
Note that S-\textsc{PROBES}-4 does \emph{not} apply at $p=20$, since
$20 \not\equiv 0,3 \pmod{12}$.

\begin{remark}
An explicit block listing for $S(2,5,25)$ --- equivalently a
BIBD$(25,5,1)$ with 30 blocks is given in Example~4 of
\citet{oruc2009bibd}, constructed via mutually orthogonal Latin squares
of order~5 (equivalently, the affine plane $\mathrm{AG}(2,5)$).
This directly gives the S-\textsc{PROBES}-5 block structure for $p=24$.
\end{remark}

\subsection{Added-Noise Variance of Sufficient-Statistic Sums}
\label{app:ss_sum_variances}

The block constructions above recover the same sufficient-statistic
\emph{sums} as full PROBES, but combine different numbers of noisy cells.
Table~\ref{tab:ss_sum_variances} compares their marginal added-noise
variances with a joint Gaussian release. Throughout this comparison,
the dataset is fixed, adjacency is add/remove, and features and response
lie in $[0,1]$. Thus the quantities describe randomness introduced by the
privacy mechanism, not sampling variability or prediction error. The
intercept and its noisy count are included. The Laplace-cell constructions
provide pure $\eps$-DP, whereas the joint Gaussian mechanism provides
approximate $(\eps,\delta)$-DP.

\paragraph{Notation and calibration.}
With $p$ features, the intercept-augmented Gram matrix and cross-product
vector contain
\[
K=\frac{(p+1)(p+4)}{2}
\]
unique entries in total. For S-PROBES, write $B$ for the number of
Steiner blocks of size $k$, $D$ for the number of disjoint diagonal
blocks of size $d$, and $M=B+D$. Here $d$ denotes the diagonal block
size, not the Bernstein order used elsewhere. Each pair of variables
occurs in one Steiner block, and each variable occurs in
$r=p/(k-1)$ such blocks. These expressions assume that the required
Steiner design and an exact diagonal partition exist: $d\mid p$ and
$D=p/d$. In the variants above, $d=k-1$, so this requires
$(k-1)\mid p$; for a different diagonal size $d$, use $M=B+p/d$.
The uniform diagonal-cell variance formulas below assume every diagonal
block has size $d$, and do not apply unchanged to a smaller final block.
Define
\[
V_G=\frac{2K}{\eps^2}\log\!\left(\frac{1.25}{\delta}\right),
\qquad v_S=2\left(\frac{M}{\eps}\right)^2.
\]
The joint vector has $\ell_2$-sensitivity $\sqrt K$; $V_G$ is its
classical sufficient Gaussian noise variance for $0<\eps\leq1$ and
$0<\delta<1$. In contrast, $v_S$ is the variance of one Laplace
cell noise at scale $M/\eps$.

\paragraph{Recovery and interpretation.}
A pair or feature--response product sums $2^{k-2}$ cells from its
unique Steiner block, while a square sums $2^{d-1}$ cells from its
diagonal block. A first moment sums $2^{k-1}$ cells per containing
Steiner block; averaging the $r$ independent block estimates divides
its variance by $r$. Similarly, averaging the $D$ diagonal-block
count estimates divides their variance by $D$. These cell counts
give the S-PROBES factors in Table~\ref{tab:ss_sum_variances}; the
full PROBES factors follow from the Bernstein recovery weights.
For fixed block sizes, the S-PROBES expressions grow polynomially
with $p$, rather than carrying the full tensor's $3^p$ factor.
However, estimates reconstructed from shared cells can have correlated
errors. Marginal variances alone therefore do not establish an ordering
of regression accuracy, and the Gaussian and Laplace columns represent
different privacy guarantees.

\begin{table}[!t]
\color{black}
\captionsetup{font={color=black},labelfont={color=black}}
\centering
\caption{Marginal added-noise variances of sufficient-statistic sums,
conditional on the dataset.}
\label{tab:ss_sum_variances}
\small
\setlength{\tabcolsep}{0.5pt}
\renewcommand{\arraystretch}{1.8}
\begin{tabular*}{\columnwidth}{@{\extracolsep{\fill}}l>{\centering\arraybackslash}m{0.42\columnwidth}cc@{}}
\toprule
Sum & Joint Gaussian & S-PROBES & Full PROBES\\
\midrule
$n$ & $\frac{(p+1)(p+4)}{\eps^2}\log(1.25/\delta)$ & $\displaystyle v_S\frac{2^d}{D}$ & $\displaystyle\frac{4(3^p)}{\eps^2}$\\
$\sum_i x_{ij}$ & $\frac{(p+1)(p+4)}{\eps^2}\log(1.25/\delta)$ & $\displaystyle v_S\frac{2^{k-1}}r$ & $\displaystyle\frac{5(3^{p-1})}{\eps^2}$\\
$\sum_i y_i$ & $\frac{(p+1)(p+4)}{\eps^2}\log(1.25/\delta)$ & $\displaystyle v_S\frac{2^{k-1}}r$ & $\displaystyle\frac{2(3^p)}{\eps^2}$\\
$\sum_i x_{ij}^2$ & $\frac{(p+1)(p+4)}{\eps^2}\log(1.25/\delta)$ & $v_S2^{d-1}$ & $\displaystyle\frac{4(3^{p-1})}{\eps^2}$\\
\makecell[l]{$\sum_i x_{ij}x_{i\ell}$\\$j\ne\ell$} & $\frac{(p+1)(p+4)}{\eps^2}\log(1.25/\delta)$ & $v_S2^{k-2}$ & $\displaystyle\frac{25(3^{p-2})}{4\eps^2}$\\
$\sum_i x_{ij}y_i$ & $\frac{(p+1)(p+4)}{\eps^2}\log(1.25/\delta)$ & $v_S2^{k-2}$ & $\displaystyle\frac{5(3^{p-1})}{2\eps^2}$\\
\bottomrule
\end{tabular*}
\end{table}

\subsection{Joint Laplace and Gaussian Baseline Algorithms}
\label{sec:naive_baselines}

Each baseline makes one joint release of the complete OLS
sufficient-statistic vector at the full privacy budget.
Assume add/remove adjacency, a fixed public choice of $p$ features,
and fixed public preprocessing that maps each record into $[0,1]^{p+1}$.
The queries below are sums, not averages; the count is privatized rather
than assumed public. For a fixed dataset, let
$z_i=(1,x_{i1},\ldots,x_{ip})^\top$, $G=\sum_i z_i z_i^\top$,
and $b=\sum_i z_i y_i$. Stack the unique upper-triangular entries
of $G$ followed by $b$ into $q\in\mathbb R^K$, with
$K=(p+1)(p+4)/2$. Every record contributes $K$ values in $[0,1]$,
so the exact joint add/remove sensitivities are
\[
 \Delta_1(q)=K\quad\text{(Laplace)},\qquad
 \Delta_2(q)=\sqrt K\quad\text{(Gaussian)}.
\]
Adding or removing the all-ones record attains both bounds on the
stated domain. Sensitivity $1$ applies to an individual scalar statistic,
not to this joint vector. In contrast, the full PROBES simplex embedding
has joint $\ell_1$-sensitivity $1$. These calculations use each unique
Gram entry once and the stated add/remove adjacency, not replacement
adjacency or a differently normalized workload.

\paragraph{Joint Laplace release.}
Release $\hat q=q+Z$ once, with independent coordinates
$Z_c\sim\Lap(\Delta_1(q)/\eps)=\Lap(K/\eps)$.
The full budget $\eps$ calibrates this single vector mechanism to its
$\ell_1$-sensitivity $K$; there is no per-statistic budget allocation.
It provides pure $\eps$-DP, with marginal
added-noise variance $2K^2/\eps^2$ as in
equation~\eqref{eq:naive_joint_variance}.

\paragraph{Joint Gaussian release.}
Release $\hat q=q+Z$ once, with
$Z\sim\mathcal N(0,\sigma_G^2 I_K)$, calibrated to the joint
$\ell_2$-sensitivity $\Delta_2(q)=\sqrt K$ at the full
$(\eps,\delta)$ budget. For $\eps>0$ and $0<\delta<1$, use analytic
Gaussian calibration: choose $\sigma_G>0$ so that, with
$a=\sqrt K/\sigma_G$,
\[
 \Phi_{\rm N}\!\left(\frac a2-\frac{\eps}{a}\right)
 -e^{\eps}\Phi_{\rm N}\!\left(-\frac a2-\frac{\eps}{a}\right)
 =\delta,
\]
where $\Phi_{\rm N}$ is the standard normal distribution function.
This one release provides $(\eps,\delta)$-DP; its marginal variance
is $\sigma_G^2$. Unlike the classical sufficient formula $V_G$ in
Section~\ref{app:ss_sum_variances}, this calibration also applies
at $\eps=10$. In Tables~\ref{tab:covertype_low} and
\ref{tab:correlated_combined}, $\delta=n^{-2}$ uses the total
training-plus-test benchmark size, treated as a fixed public parameter;
it is not computed from the noisy count.

\paragraph{Variance comparison of joint Gaussian and S-PROBES at $p=60$.}
Set $\eps=1$, $n=10{,}000p=600{,}000$ as a fixed public size
parameter, and $\delta=n^{-2}\approx2.77778\times10^{-12}$.
For S-\textsc{PROBES}-5 with diagonal groups of size $d=4$,
$p=60$ satisfies the design condition $p\equiv0,4\pmod{20}$.
There are $B=183$ Steiner blocks and $D=15$ diagonal blocks, so
$M=198$, $r=p/4=15$, and $v_S=2M^2=78{,}408$.
Using Table~\ref{tab:ss_sum_variances}, the count and each
first-moment sum have variance
\[
 v_S\frac{16}{15}=83{,}635.2,
\]
whereas each square, feature--feature cross-product, and
feature--response cross-product sum has variance
\[
 v_S\,8=627{,}264.
\]
For the joint Gaussian release, $K=(p+1)(p+4)/2=1{,}952$; solving the analytic
Gaussian calibration equation above gives
$\sigma_G^2\approx80{,}212.3$ for every released statistic.
Thus the Gaussian release has lower marginal variance for every sufficient-statistic
sum in this example. These are analytical added-noise variances for
unnormalised sums, not measured regression errors, and use analytic
calibration rather than the classical sufficient variance $V_G$.
The comparison allows approximate $(\eps,\delta)$-DP for the Gaussian release,
whereas S-\textsc{PROBES}-5 provides pure $\eps$-DP.

\paragraph{Reconstruction and stabilization.}
Unpack the perturbed vector into $\hat G$ and $\hat b$, copying
each noisy upper-triangular Gram entry to its symmetric position
without adding a second noise draw. Use a spectral shift with
noise-standard-deviation ridge $\rho_L=\sqrt2 K/\eps$ for
the Laplace release and $\rho_G=\sigma_G$ for the Gaussian release. Algorithm~\ref{alg:naive_baselines}
summarizes the release and solve. The shift ensures a positive-definite
matrix for every noise realization; it is a stabilization heuristic,
not an optimal-accuracy guarantee. Reconstruction, stabilization and
prediction are post-processing of the private vector. The stated
guarantees assume public bounds and preprocessing; data-dependent
feature selection or normalization requires separate privacy accounting.

\begin{algorithm}[t]
\color{black}
\SetAlCapFnt{\color{black}}
\SetAlCapNameFnt{\color{black}}
\SetAlFnt{\small}
\SetNlSty{red}{}{}
\caption{Joint Laplace / Gaussian OLS Sufficient-Statistic Release}
\label{alg:naive_baselines}
\SetKwInOut{Input}{Input}
\SetKwInOut{Output}{Output}
\Input{Records in $[0,1]^{p+1}$ under fixed public preprocessing;
add/remove adjacency; public $p$, $\eps>0$; mode L or G;
public $0<\delta<1$ for mode G}
\Output{Private regression coefficients $\hat\theta$}
$K\gets(p+1)(p+4)/2$; $z_i\gets(1,x_{i1},\ldots,x_{ip})^\top$\;
$G\gets\sum_i z_i z_i^\top$; $b\gets\sum_i z_i y_i$\;
$q\gets$ upper triangle of $G$, followed by $b$\;
$\Delta_1\gets K$; $\Delta_2\gets\sqrt K$\;
\eIf{mode is L}{
  $Z_c\overset{\mathrm{iid}}{\sim}\Lap(\Delta_1/\eps)$, $c=1,\ldots,K$\;
  $\rho\gets\sqrt2 K/\eps$\;
}{
  Find $a>0$ numerically by bisection such that\;
  $\Phi_{\rm N}(a/2-\eps/a)
  -e^{\eps}\Phi_{\rm N}(-a/2-\eps/a)=\delta$\;
  $\Phi_{\rm N}$ is the standard normal CDF\;
  $\sigma_G\gets\Delta_2/a=\sqrt K/a$\;
  $U_c\overset{\mathrm{iid}}{\sim}\mathcal N(0,1)$,
  $c=1,\ldots,K$\;
  $Z_c\gets\sigma_G U_c$, $c=1,\ldots,K$\;
  $\rho\gets\sigma_G$\;
}
Form the single private vector $\hat q\gets q+Z$\;
Unpack $\hat q$ into symmetric $\hat G$ and $\hat b$\;
$s\gets\rho+\max\{0,-\lambda_{\min}(\hat G)\}$\;
\Return{$\hat\theta=(\hat G+sI_{p+1})^{-1}\hat b$}\;
\end{algorithm}

\endgroup

\begingroup\color{black}
\section{Coefficient Error and Inverse-Amplification Bounds}
\label{sec:error_revision}

All statements below condition on a fixed dataset of $n$ records with
$p$ features and $x_{ij},y_i\in[0,1]$. The design matrix
$X\in\mathbb R^{n\times(p+1)}$ has rows $(1,x_{i1},\ldots,x_{ip})$.
We use $\|\cdot\|_2$, $\|\cdot\|_{\rm op}$, and $\|\cdot\|_F$
for the Euclidean vector norm, its induced matrix operator norm, and
the Frobenius norm, respectively. Write $G=X^\top X$,
$b=X^\top y$, $\theta=G^{-1}b$, and assume
$\lambda_{\min}(G/n)\geq\kappa>0$. Let $P_X=XG^{-1}X^\top$
be the orthogonal projector onto the column space of $X$. Then
$\|\theta\|_2\leq\kappa^{-1/2}$, since
$n\kappa\|\theta\|_2^2\leq\theta^\top G\theta
=\|P_Xy\|_2^2\leq\|y\|_2^2\leq n$.
Let $\hat A$ and $\hat{\bm b}$ be the recovered noisy Gram matrix
and right-hand side in Algorithm~\ref{alg:rPROBES}. Define
$E=\hat A-G\in\mathbb R^{(p+1)\times(p+1)}$ and
$e=\hat{\bm b}-b\in\mathbb R^{p+1}$; $E$ is symmetric.
With $[a]_+=\max\{a,0\}$ and $I=I_{p+1}$ the identity matrix, define
\begin{align*}
 s&=\rho+[-\lambda_{\min}(G+E)]_+,\qquad \rho>0,\\
 A&=G+E+sI,\qquad \hat\theta=A^{-1}(b+e).
\end{align*}
The shift ensures $\lambda_{\min}(A)\geq\rho$, so the stabilized
solve is always defined.

\begin{lemma}[Deterministic perturbation bound]
\label{lem:perturbation_revision}
If $\|E\|_{\rm op}\leq\eta$ and
$\|e-E\theta\|_2\leq\zeta$, then $n\kappa>\eta$ implies $s=\rho$ and
\begin{equation}
 \|\hat\theta-\theta\|_2\leq
 \frac{\zeta+\rho/\sqrt\kappa}{n\kappa+\rho-\eta}.
 \label{eq:error_revision}
\end{equation}
Absolute contraction $\|A^{-1}\|_{\rm op}\leq1$ holds when
$n\kappa+\rho-\eta\geq1$. For any $\tau>0$, relative amplification
$\|A^{-1}\|_{\rm op}/\|G^{-1}\|_{\rm op}\leq1+\tau$ follows if
$n\kappa\geq(1+\tau)(\eta-\rho)_+/\tau$.
\end{lemma}
\begin{proof}
Weyl's inequality gives $\lambda_{\min}(G+E)\geq n\kappa-\eta>0$.
The identity $A(\hat\theta-\theta)=e-E\theta-\rho\theta$ proves
the error bound. For the ratio, put $g=\lambda_{\min}(G)\geq n\kappa$.
Then $\|A^{-1}\|_{\rm op}/\|G^{-1}\|_{\rm op}\leq g/(g+\rho-\eta)$.
If $\eta\leq\rho$ this is at most one; otherwise the stated
condition makes it at most $1+\tau$. The absolute claim follows
from the same lower bound on $\lambda_{\min}(A)$.
\end{proof}

\begin{theorem}[Full-tensor high-probability bound]
\label{thm:full_error_revision}
Let $p\geq1$, $\eps>0$, $0<\alpha<1$, and
$u=\log(2(p+1)(p+2)/\alpha)$, where $\log$ is the natural logarithm.
For the order-$2$ full PROBES release, define
\begin{align*}
 \eta&=\frac{\sqrt{3^{p-2}(25p^2+143p+144)u/2}+(p+1)u}{\eps},\\
 q&=\frac{\sqrt{3^{p-1}(5p+12)u}+\sqrt{p+1}\,u}{\eps},
 \qquad \zeta=q+\eta/\sqrt\kappa.
\end{align*}
With probability at least $1-\alpha$ over the joint release noise,
$\|E\|_{\rm op}\leq\eta$ and $\|e-E\theta\|_2\leq\zeta$.
If also $n\kappa>\eta$, the coefficient-error bound of
Lemma~\ref{lem:perturbation_revision} holds on this same event;
its inverse-norm bounds hold under their stated conditions.
No independence between $E$ and $e$ is required.
\end{theorem}
\begin{proof}
Index the $2\cdot3^p$ cells by
$c=(\bm k,r)\in\{0,1,2\}^p\times\{y,\bar y\}$.
For each cell, define the symmetric recovery matrix
$U_c\in\mathbb R^{(p+1)\times(p+1)}$ and vector
$w_c\in\mathbb R^{p+1}$ by
\begin{align*}
 (U_c)_{00}&=1,\\
 (U_c)_{0j}=(U_c)_{j0}&=k_j/2,\\
 (U_c)_{jj}&=\mathbf{1}_{\{k_j=2\}},\\
 (U_c)_{j\ell}&=k_jk_\ell/4\quad(j\ne\ell),\\
 w_c&=\mathbf{1}_{\{r=y\}}(1,k_1/2,\ldots,k_p/2)^\top,
\end{align*}
where $1\leq j,\ell\leq p$ and $\mathbf{1}_{\{\cdot\}}$ is the
indicator, equal to one when its condition holds and zero otherwise.
These are the order-$2$ monomial recovery coefficients in
Algorithm~\ref{alg:rPROBES}. Thus, for the independent cell noises
$\xi_c\sim\Lap(1/\eps)$,
\[
 E=\sum_c\xi_c U_c,\qquad e=\sum_c\xi_c w_c.
\]
Put $d=p+1$. The monomial identities give
\begin{align*}
 \sum_c\|U_c\|_F^2&=3^{p-2}(25p^2+143p+144)/8,\\
 \sum_c\|w_c\|_2^2&=3^{p-1}(5p+12)/4.
\end{align*}
Explicitly, the count weights have squared sum $2\cdot3^p$;
each first-moment weight vector has squared sum
$2(5/4)3^{p-1}$, each square has $2\cdot3^{p-1}$, and each
feature-pair weight vector has $2(25/16)3^{p-2}$.
Thus the Gram sum is
\begin{align*}
 &2\cdot3^p+4p(5/4)3^{p-1}+2p3^{p-1}\\
 &\hspace{20pt}+2p(p-1)(25/16)3^{p-2},
\end{align*}
which simplifies to the first formula. The response-sum and
feature-response vectors give
$3^p+p(5/4)3^{p-1}$, proving the second formula.
Each $U_c$ has
entries in $[0,1]$, so $\|U_c\|_{\rm op}\leq d$; similarly
$\|w_c\|_2\leq\sqrt d$.

For completeness, let $V_c\in\mathbb R^{m\times m}$ be a finite
family of fixed symmetric matrices, and let $\xi_c\sim\Lap(a)$
be independent with $a>0$. For any $u>0$, the self-adjoint
Laplace series $H=\sum_c\xi_c V_c$ satisfies
\[
 \Pr\{\|H\|_{\rm op}>\sqrt{2Vu}+Lu\}\leq2m e^{-u},
\]
where $m$ is the matrix dimension,
$V=2a^2\|\sum_c V_c^2\|_{\rm op}$ and
$L=a\max_c\|V_c\|_{\rm op}$; any larger $V,L$ remain valid.
To see this, functional calculus gives
$\mathbb E e^{t\xi_cV_c}=(I_m-a^2t^2V_c^2)^{-1}$ for $0<t<1/L$,
where $I_m$ is the $m\times m$ identity matrix.
The scalar inequality $-\log(1-x^2)\leq x^2/(1-x)$, for $0\leq x<1$,
implies the matrix log-mgf bound
$\log\mathbb E e^{t\xi_cV_c}\preceq
a^2t^2V_c^2/(1-tL)$, where $\preceq$ denotes the
positive-semidefinite order. The independent matrix Laplace-transform
inequality~\citep[Theorem 3.6]{tropp2012user} therefore gives
\[
 \Pr\{\lambda_{\max}(H)>x\}
 \leq m\inf_{0<t<1/L}
 \exp\!\left(-tx+\frac{Vt^2}{2(1-Lt)}\right).
\]
Indeed the sum of the log-mgfs is at most
$t^2(2a^2\sum_c V_c^2)/(2(1-Lt))$, whose largest eigenvalue
is at most $Vt^2/(2(1-Lt))$; its exponential has trace at most
$m$ times the exponential of that number.
Set $x=\sqrt{2Vu}+Lu$ and
$t=\sqrt{2u/V}/(1+L\sqrt{2u/V})$. Substitution makes the
exponent exactly $-u$. Apply the same argument to $-H$ and add
the two tails. The degenerate case $V=0$ has $H=0$ almost surely.
This establishes the displayed Laplace-series inequality.

Apply this with $a=1/\eps$ to $U_c$, bounding
$\|\sum_c U_c^2\|_{\rm op}$ by
$\sum\|U_c\|_F^2$. For the vector use self-adjoint dilations

\mbox{
$\left[\begin{smallmatrix}0&w_c^\top\\w_c&0\end{smallmatrix}\right]$
}
and bound their variance proxy by $2\sum\|w_c\|_2^2/\eps^2$.
The failure probabilities are at most $2d e^{-u}$ and
$2(d+1)e^{-u}$. Their sum is at most $\alpha$ because
$4d+2\leq2d(d+1)$ for $d\geq2$.
Finally $\|e-E\theta\|_2\leq\|e\|_2
+\|E\|_{\rm op}\|\theta\|_2\leq q+\eta/\sqrt\kappa$.
\end{proof}

\begin{corollary}[Sufficient sample size]
\label{cor:samples_revision}
Under the assumptions of Theorem~\ref{thm:full_error_revision}, set
$\rho=2\cdot3^{(p-1)/2}/\eps$. For $0<t\leq1$, $\tau>0$,
the sufficient condition
\[
 n>\frac{1}{\kappa}+
 \frac{12p(3^{p/2}\sqrt u+u)}{\kappa\eps}
 \max\left\{\frac{1}{t\sqrt\kappa},\frac{1}{\tau}\right\}
\]
ensures coefficient error at most $t$, relative inverse amplification
at most $1+\tau$, and absolute inverse norm at most one, jointly
with probability at least $1-\alpha$.
\end{corollary}
\begin{proof}
The intercept implies $\kappa\leq1$. Put
$H=12p(3^{p/2}\sqrt u+u)/\eps$ and $a_0=3^{p/2}\sqrt u/\eps$.
Directly from the formulas, $\eta\leq(21/5)pa_0+2pu/\eps$,
$q\leq(12/5)pa_0+2pu/\eps$, and $\rho\leq(6/5)pa_0$;
the last inequality uses $u>1$. Thus
$2\eta+q+\rho\leq H$ and $\eta\leq H/2$.
The condition implies $n\kappa>1+H/(t\sqrt\kappa)$, so
the denominator in~\eqref{eq:error_revision} exceeds one and
\[
 t(n\kappa+\rho-\eta)>
 \frac{H}{\sqrt\kappa}+t\rho-t\eta
 \geq q+\frac{\eta+\rho}{\sqrt\kappa}.
\]
Here $t\leq1$, $\sqrt\kappa\leq1$, and
$H\geq2\eta+q+\rho$ suffice for the last inequality.
For relative amplification, the sample condition also gives
$n\kappa>H\max\{1,1/\tau\}\geq(1+1/\tau)\eta$,
which is at least the sufficient threshold in the lemma.
\end{proof}

\endgroup

\end{document}